\documentclass[10pt, journal]{IEEEtran}

\usepackage[english]{babel}
\usepackage{blindtext}
\usepackage[framemethod=TikZ]{mdframed}

\usepackage{lipsum}
\usepackage{epsfig,amsmath,url,bm}
\usepackage{amsthm}
\usepackage{caption}
\usepackage{subcaption}
\usepackage{epsf,algorithm,algorithmic}
\usepackage{balance}

\DeclareMathOperator*{\argmin}{argmin}

\title{Tomography Based Learning for Load Distribution through Opaque Networks}
\begin{document}

\author{Shenghe Xu,\textsuperscript{\rm 1}
Murali Kodialam,\textsuperscript{\rm 2}
T.V. Lakshman,\textsuperscript{\rm 2}
Shivendra S. Panwar,\textsuperscript{\rm 1}\\
\textsuperscript{\rm 1}Department of Electrical and Computer Engineering, NYU Tandon School of Engineering, Brooklyn, NY \\
\textsuperscript{\rm 2}Nokia Bell Labs, Crawford Hill, NJ\\
shenghexu@nyu.edu,
\{murali.kodialam, tv.lakshman\}@nokia-bell-labs.com, panwar@nyu.edu
}

\maketitle

\begin{abstract}
Applications such as virtual reality and online gaming require low delays for acceptable user experience. A key task for over-the-top (OTT) service providers who provide these applications is sending traffic through the networks to minimize delays. OTT traffic is typically generated from multiple data centers which are multi-homed to several network ingresses. However, information about the path characteristics of the underlying network from the ingresses to destinations is not explicitly available to OTT services. These can only be inferred from external probing. 
In this paper, we combine network tomography with machine learning to minimize delays. We consider this problem in a general setting where traffic sources can choose a set of ingresses through which their traffic enter a black box network. 
The problem in this setting can be viewed as a reinforcement learning problem with constraints on a continuous action space, which to the best of our knowledge have not been investigated by the machine learning community. Key technical challenges to solving this problem include the high dimensionality of the problem and handling constraints that are intrinsic to networks. Evaluation results show that our methods achieve up to 60\% delay reductions in comparison to standard heuristics. Moreover, the methods we develop can be used in a centralized manner or in a distributed manner by multiple independent agents.
\end{abstract}


\newcommand{\squeezeup}{\vspace{-2.5mm}}
\section{Introduction}
Recent emerging applications including virtual reality, online or cloud gaming require low delay for acceptable user experience \cite{albert2017latency, claypool2010latency}. 
Minimizing delay by optimizing load distribution through underlying networks is an important task for providers of these services. 
However, since these services are often ``over-the-top" services, the providers do not have full knowledge of the underlying networks and have to make load distribution decisions based purely on inference of the network characteristics from edge-based observations. Inferring network characteristics from external observations, called ``network tomography", has been extensively studied. Early work in network tomography focused on the ``inverse problem" of estimating traffic matrices from link-level observations only \cite{vardi1996network, zhang2003fast, soule2007estimating}. In this paper, our interest is in "active tomography" where probes from the network periphery are used to infer internal network characteristics \cite{caceres1999multicast, chen2003tomography, duffield2006network, bu2002network}. 

We view the network as a black box with most of the network's features of interest for load distribution purposes being not directly observable. Most of the important information for performance optimization is hidden and hard to measure. For example, without information from the Internet Service Providers (ISPs), inferring the routing structure of the network is often an impossible task. While ping and traceroute may provide some insight, routers may not respond to these kinds of probe packets. 

\begin{figure}[t]
\centering
\includegraphics[width=.48\textwidth]{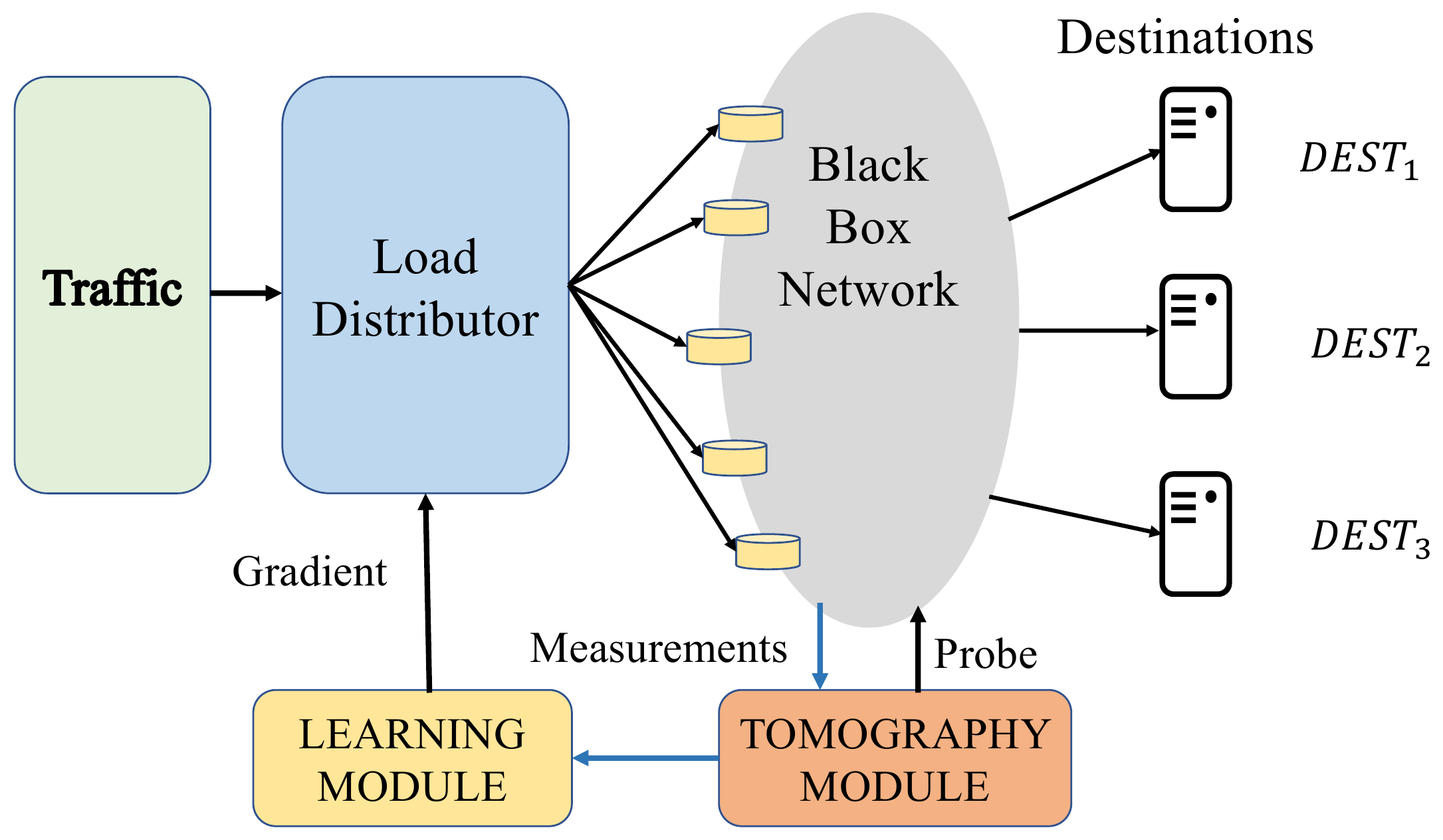}
\caption{Sending traffic through a black box network}
\label{Fig: Fig5_crop}
\end{figure}
We consider the scenario in Figure \ref{Fig: Fig5_crop} where there are a set of sources that have traffic to send through the black box network to a set of destinations.The traffic sources know the ingresses into the network and can send probes though the network to different endpoints. The probes are echoed by the endpoints; returning probes can be observed to determine the network's response to probing and infer behavior of the underlying network. 
Sources have the choice of distributing their traffic over the set of ingresses to the network -- a source may send all its traffic to a destination through a particular ingress or it may decide to distribute its traffic to the different ingresses in proportions that minimize the total delay through the network. Because the network is a black box, the information that is needed to make the optimal distribution choice can only be gleaned from external observations of responses to past actions. For the load-distribution problem, we use the history of tomography-obtained responses to past actions to train a neural network which we then combine with reinforcement learning to make future load distribution decisions that minimize delay through the underlying black box network. 
Unlike network tomography, where the primary goal is network monitoring and measurement, our goal here is automated performance optimization where information obtained through tomography is used to optimize performance through black box networks. To this end, we combine learning-based methods with network tomography to optimize performance through black box networks. 
The scenario we consider of a black box network through which traffic has to be distributed to minimize delay, is representative of many important use cases in networking. A few of these are outlined below:
\begin{itemize} 
\begin{figure}[t]
\centering
\includegraphics[width=.48\textwidth]{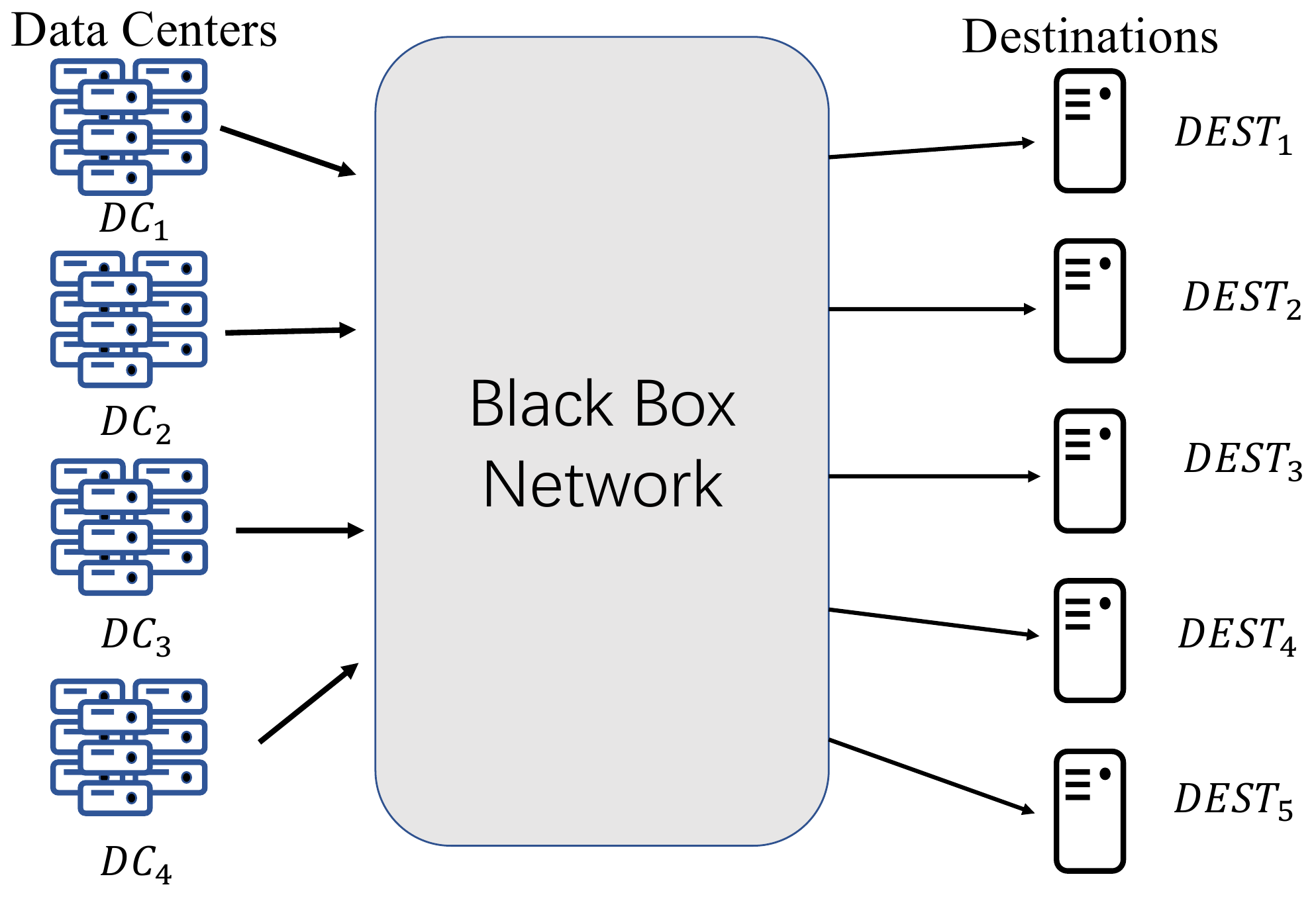}
\caption{Load distribution from data centers to black box network}
\label{Fig: DC}
\end{figure}
\item A content provider with content replicated in multiple data centers (Figure \ref{Fig: DC}) has to decide what fractions of requested traffic have to be drawn from the different data centers and consequently how this traffic is to be distributed to the different ingresses of the network to which the data centers are connected. To the content provider, the network characteristics are not directly observable and so the load distribution decision has to be based on network characteristics observable from the network edge. 
\begin{figure}[!t]
\centering
\includegraphics[width=.49\textwidth]{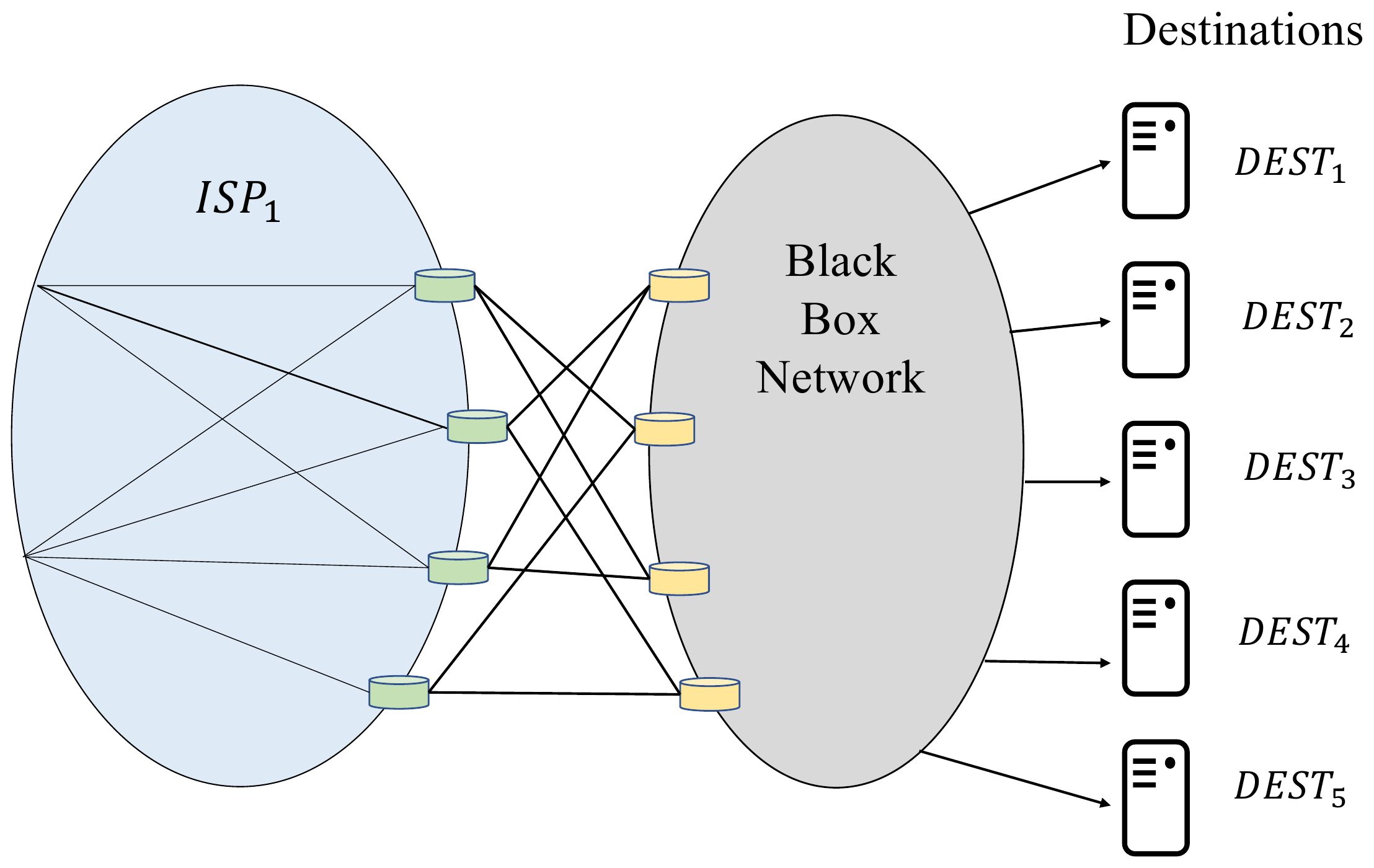}
\caption{Egress picking to minimize delays}
\label{Fig: EP}
\end{figure}
\item An ISP has to decide how traffic towards downstream destinations has to be split amongst multiple egresses from its network into downstream networks (Figure \ref{Fig: EP}), i.e., the ISP has to pick the optimal split of traffic to the different ingresses of downstream networks. Since downstream networks may belong to different providers (and hence different Autonomous Systems), the internals of downstream networks are not directly observable. Moreover, BGP does not provide path metric information sufficient for fine-grained performance optimization. Hence, traffic distribution decisions to downstream ISPs have to be based on tomography-based information obtained by probing through the black box downstream networks. 
\begin{figure}[t]
\centering
\includegraphics[width=.49\textwidth]{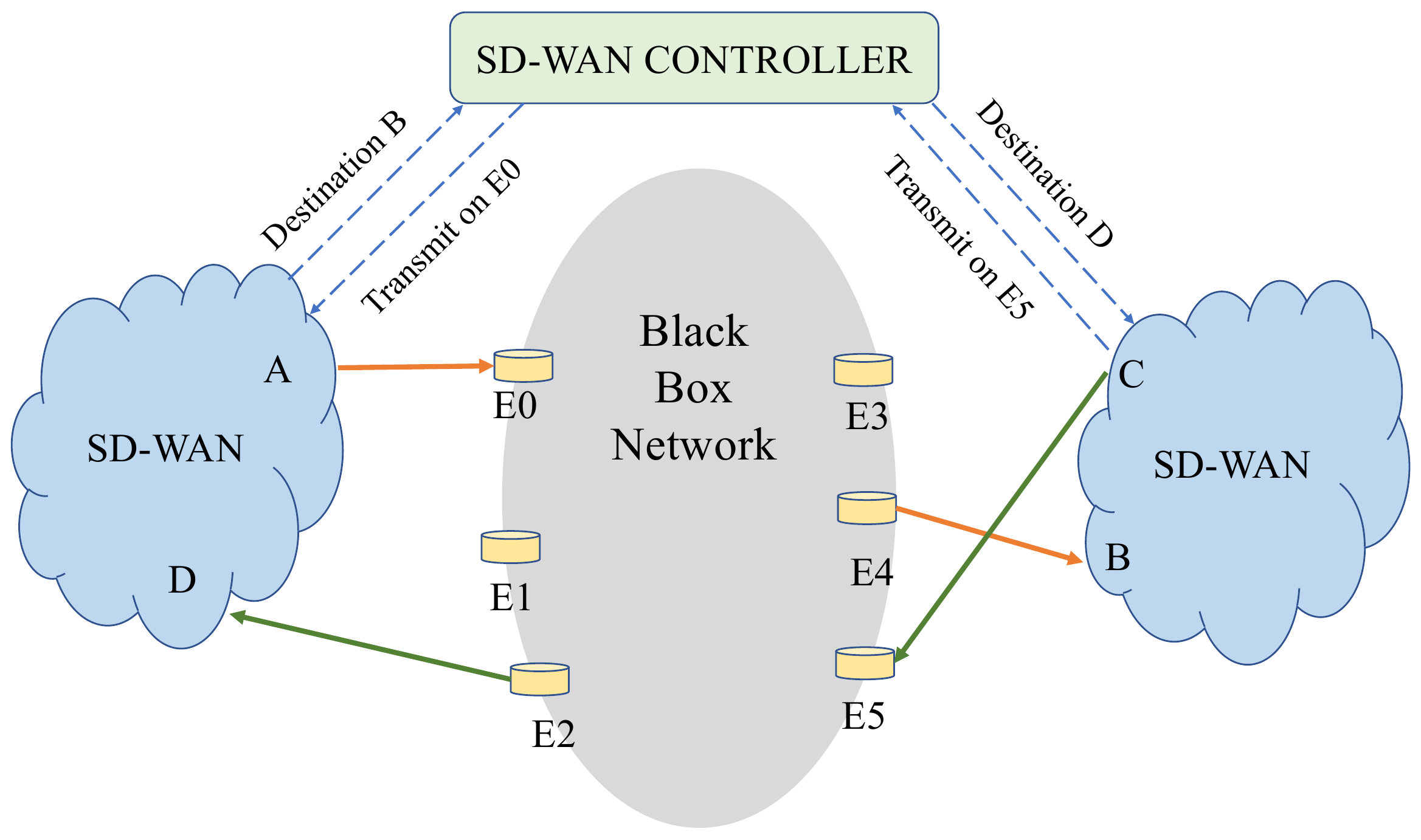}
\caption{Load distribution from SD-WAN gateway nodes to underlay networks}
\label{Fig: SDWAN}
\end{figure}
\item In Software-Defined WANs (SD-WANs), gateway nodes can be multi-homed and have to decide how to split traffic to different underlay ingresses to minimize delays through the underlay network. This is shown in Figure \ref{Fig: SDWAN}. The underlay is a black box for the SD-WAN nodes and the only information about the underlay available to the SD-WAN nodes is by network tomography
\begin{figure}[t]
\centering
\includegraphics[width=.49\textwidth]{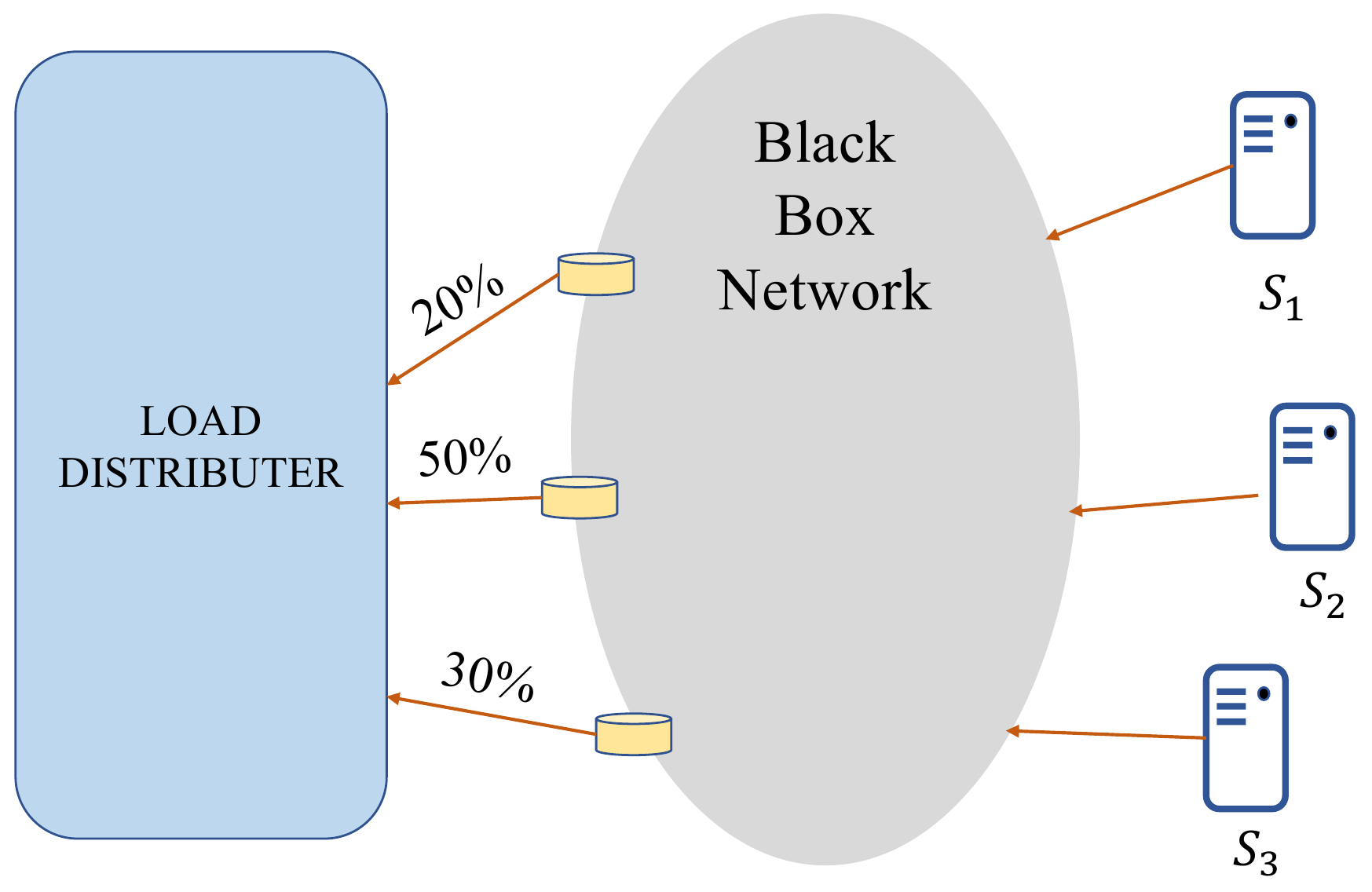}
\caption{Network load balancing over black box networks}
\label{Fig: LB}
\end{figure}
\item Network load balancers \cite{handigol2010aster} that distribute incoming demands to a set of distributed servers, as shown in Figure \ref{Fig: LB}, can view the combination of the underlying network and servers as a black box and can optimize the load distribution based on the observed delays in response to past actions. 

\item Segment routing \cite{filsfils2015segment} has been proposed for traffic engineering in networks to improve service quality and avoid link congestion. In segment routing, the end-to-end path is composed of segments (as in Figure \ref{Fig: SR_S}) where the end-points of the segments are carefully chosen to avoid network congestion. If segment routing is done at the overlay layer, the segments themselves are routed by IGP picked paths in the the underlay. The optimal choice of segment end-points and the optimal split of traffic through different segment routed paths to the same destination will need to be done based on information available at the sources through network tomography.
\end{itemize}

In this paper, we focus on the scenario that the underlay network is acting as a complete black box. We assume that the only information that we can obtain from the black box is via 
end-to-end tomography measurement of average delays. 
We use these measurements in conjunction with  past decisions that resulted in the 
observed delays to make current decisions. The decisions made have to  be robust 
to changing underlying network conditions and be responsive to changes in network topology like link or node failures. 

Though tomography based techniques have been used to identify hot spots in networks (See for example, \cite{guo2015pingmesh}), it is challenging to use it in a machine learning 
based approach to optimize load distribution. This is due to the fact that the 
decision space (how to split the traffic) as well as the rewards (tomography measurements) are continuous and high dimensional. This makes it very difficult to use 
traditional reinforcement learning based techniques to solve the problem. However, recent advances in reinforcement learning, especially actor-critic networks, make it possible to 
implicitly store the actions and rewards in a neural network. Our problem, apart from being continuous and high dimensional, also has constraints on the set of actions. Instead of 
actor-critic learning, we propose a critic only learning algorithm and use 
a Frank-Wolfe \cite{frank1956algorithm} based technique to enforce the constraints. This approach leads to 
robust learning algorithms with rapid convergence. Before we outline the machine learning approach used in Section \ref{methodology}, we discuss two representative problems in more detail.

To our knowledge, the problem of reinforcement learning with constrained continuous action space has never been investigated in the machine learning field. Previous papers \cite{achiam2017constrained, dalal2018safe} included constraints on the performance during exploration for the purpose of safety, instead of adding constraints explicitly on the action space. In addition, this is the first paper that proposes to use gradient estimates provided by a neural network in a Frank-Wolfe based technique, for the purpose of solving reinforcement learning problems. 

Overall, the main contribution of this paper are
\begin{itemize}
    \item We provide a uniform model for problems including load distribution, egress picking, load balancing and segment routing. 
    \item The problems can be formulated as a reinforcement learning problem, with constraints on a continuous action space. To the best of our knowledge no prior work has proposed any method to enforce constraints on a continuous action space. 
    \item We propose the method of critic only reinforcement learning and combine it with the Frank-Wolfe method. The overall algorithm achieves better performance compared with the state-of-the-art method DDPG \cite{lillicrap2015continuous}, with higher data efficiency. 
    \item The proposed method can be used in a centralized manner, or independently by multiple distributed agents.  
\end{itemize}

\section{Related Work}
Traffic engineering problems have been extensively investigated  
\cite{feldmann2000netscope,fortz2002traffic}. In general, traffic engineering assumes that the network topology, link capacities as well as the estimated point-to-point traffic is known and the objective is to determine how to route the incident traffic 
in a congestion-free manner. While this is an appropriate model for an ISP that is 
designing MPLS tunnels or OSPF weights, full knowledge of the topology and routing cannot be assumed for "over-the-top" routing. This is the reason we have to use tomography to 
implicitly infer the topology and capacities  of the opaque network. 

Network tomography, as originally proposed in \cite{vardi1996network}, was aimed at estimating 
the traffic matrix from link measurements \cite{ zhang2003fast, soule2007estimating}. Tomography has evolved to the problem of inferring internal information of a network from end point measurements. A maximum-likelihood estimator for loss rates on internal links based on losses observed by multicast receivers was proposed in \cite{caceres1999multicast,bu2002network,chen2003tomography}. In \cite{duffield2006network} it was found that sending stripes of probe packets helps increase estimation accuracy of link loss. There has also been recent interest  \cite{guo2015pingmesh} in using large scale end-to-end pings for network diagnostics in very large networks.
In \cite{schlinker2017engineering}, performance and capacity aware routing methods were proposed to help large content providers avoid congested edges and improve user experience. A traffic controller received real time traffic and performance measurements to make routing and traffic balancing decisions. 
\cite{pujol2019steering} investigated the problem of steering large scale traffic at the ISP level. They show that traffic on long-haul links can be reduced by 30 percent if suitable egress points are recommended to a large content provider.  
The tomography literature has focused mainly on determining 
hot spots in networks by making edge-to-edge measurements. In this paper, we extend this 
idea and use tomography data to actually optimize network performance. 
Using tomography measurements for optimizing network performance is possible due to 
recent developments in the machine learning literature. The idea of using reinforcement 
learning for routing was initiated in \cite{boyan1994packet} before the recent developments in machine learning. More recently, \cite{valadarsky2017learning} investigated the problem of using machine learning for routing. These machine learning approaches assume that the network topology is known to the learning algorithm and do not 
deal with optimizing routing over  opaque networks. 
There has been recent work on using machine learning for flow scheduling \cite{chen2018auto}, congestion control \cite{winstein2013tcp, li2018qtcp,xiao2019tcp, jay2019deep} and 
optimization in video streaming\cite{mao2017neural}.
To our knowledge, this is the first work that uses the newly developed 
actor-critic reinforcement techniques for optimizing load distribution using tomographic information only.
\section{Two Representative Problems}
Though the idea of tomography based learning is applicable to several networking 
scenarios, this paper focuses on two applications, Egress Picking and Traffic Engineering using Segment Routing, to illustrate the applicability of the method. We now describe these two problems in more detail. In order to achieve better performance and scalability, network operators 
frequently split traffic across different components of a network. This is 
especially important when the capacity is asymmetric either due to 
different types of equipment deployed in different parts of the network or
due to asymmetric sharing of capacity between multiple users.
A common (implicit or explicit) objective when sharing is to minimize the average packet delay. Minimizing average packet delay leads the traffic being split roughly in proportion to the capacity. Traffic splitting is complicated by the fact that different network components are shared between multiple users and the available capacity for a given user varies over time. Both the problems that we 
study in detail are traffic splitting problems over networks where we do not have visibility into the topology or interfering traffic. However, we can use tomography to 
obtain delay estimates and we use these measurements to guide the traffic splitting 
process.  
\subsection{Egress Picking}
\begin{figure}[!t]
	\centering
	\includegraphics[width=.43\textwidth]{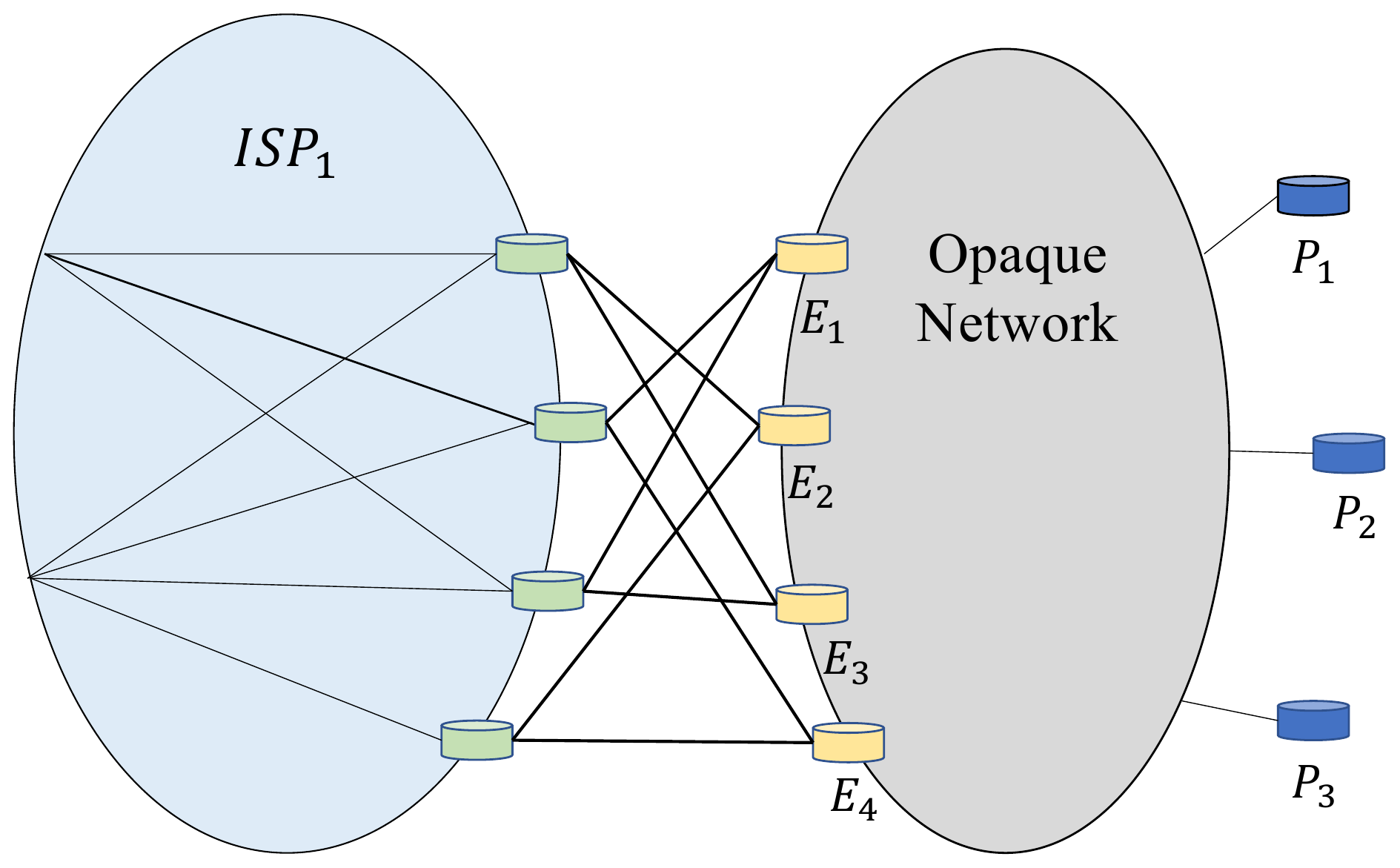}
\caption{An illustration of egress picking}
\label{Fig: EP_S}
\end{figure}

Consider an ISP that is routing traffic to destinations downstream through other 
Autonomous Systems which are opaque to the ISP originating traffic. In general,
the originating ISP has multiple choices through which it can transit traffic. The
egress picking problem \cite{kang2015efficient} is one where the originating ISP has to determine how to 
split traffic among the different egress choices in order to efficiently use the 
capacity downstream. In particular, it is important to ensure that the amount of flow sent on any path does not exceed the capacity of the components on that path, Since the capacities are not observable, it is possible to estimate whether capacities are being 
violated by measuring the delay. One way of ensuring that capacity is used efficiently is for the egress picking algorithm to minimize the mean delay. Traffic from an ISP is routed using destination 
prefixes. Each prefix is routed to a destination along an egress point. In this work, we assume that
the prefixes can be split across multiple egresses. The splitting is done such that individual flows are routed along the same path in order to avoid out of sequence packets. 
In general, there can be thousands of 
prefixes that are routed but typically there are a small subset of prefixes that carry the bulk of the traffic \cite{kohler2002observed}. Therefore, we focus attention on the top few prefixes. 
We illustrate the egress picking problem in 
 Figure \ref{Fig: EP_S}, which shows an example with four egress points and three prefixes. 
Assume that there are $n$ large destination prefixes and $m$ egress point choices.
Let $t_i$ denote the amount of traffic for prefix $i$ and let $\alpha_{ij}$ represent the fraction of traffic from prefix $i$ that is routed through egress $j$. Traffic that is routed to egress $j$ is now forwarded along some path to the destination through the opaque network. Let $D_{ij}$ represent the delay from egress $j$ to the destination of 
prefix $i$. {\em This delay $D_{ij}$ is a non-linear function of the traffic on the links of the 
opaque destination network.} The objective is to determine the split of each prefix to 
minimize the mean delay. 
\begin{equation}
    \min_{\bm{\alpha}} \frac{1}{\sum_{i=1}^n t_i}\sum_{i=1}^{n} \sum_{j=1}^{m} t_i D_{ij} \alpha_{ij} 
\end{equation}
subject to
\begin{eqnarray*}
    \sum_{j=1}^{m} \alpha_{ij} & = & 1, \forall i  \\
    \alpha_{ij} & \geq & 0 , \forall i \;\; \forall j 
\end{eqnarray*}
The $\sum_{j=1}^{m} \alpha_{ij} =  1$ constraints are called the simplex constraints. 
In addition to the simplex and non-negativity constraints there may be additional constraints that may result from 
policy considerations. For example, there may be upper bounds on the total amount of traffic that 
can be routed to a particular egress point. If this is the capacity of the egress point then the 
objective of minimizing mean delay will automatically enforce the constraint, but if it is a policy constraint, then it has to be explicitly enforced by the optimization algorithm. 
Note that the delay $D_{ij}$ is a non-linear function of $\alpha_{ij}$, and if this function is known (and convex), then we can use projected gradient descent based techniques to solve this problem. Since the network is opaque, we do not know the function $D_{ij}$ and therefore we use a tomography based learning algorithm to solve this problem.
The tomography module measures delays across the opaque network using probing and 
this is used as a feedback for the optimization algorithm.
\subsection{Traffic Engineering using Segment Routing}
\begin{figure}[!t]
	\centering
	\includegraphics[width=.49\textwidth]{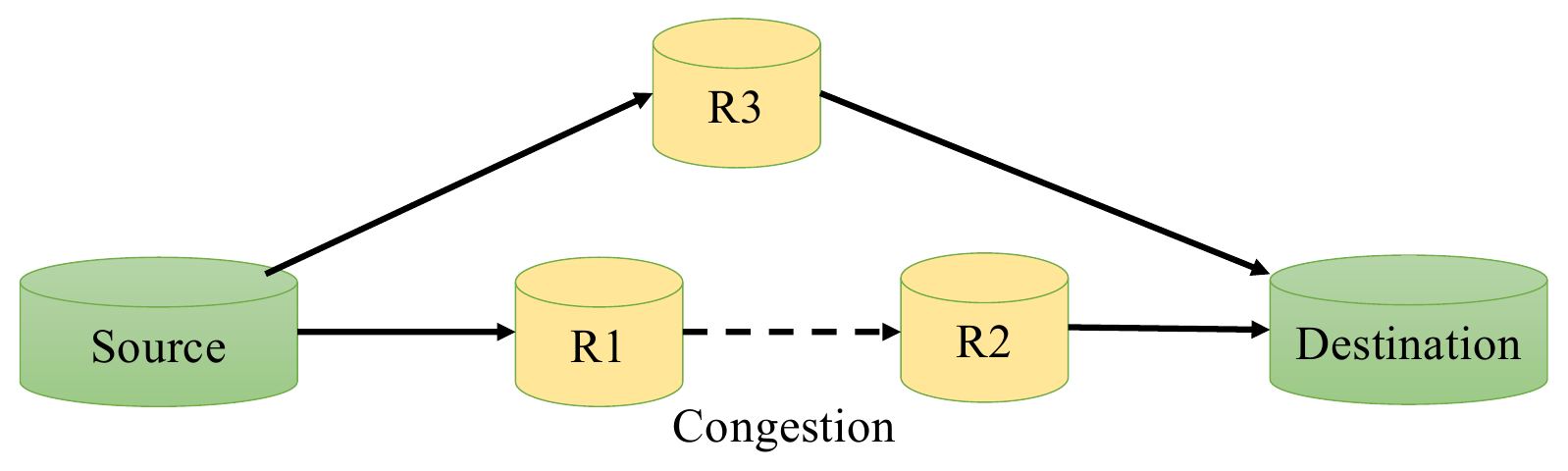}
\caption{An illustration of segment routing for congestion avoidance}
\label{Fig: SR_S}
\end{figure}

Another application of tomography based learning is traffic engineering using Segment Routing.
Segment routing is an IETF protocol for traffic engineering \cite{filsfils2015segment}. The key idea of segment routing is to break the route of a flow into several segments. Each segment is a shortest path between the two end points of the segment. Segment information is carried in the packet header and therefore there is no per-flow state maintained in the network. Assume that we have a opaque network where we only have access to a set of edge nodes. We want to route traffic between the edge nodes. 
Assume that the amount of traffic between edge node $i$ and edge node $j$ is $T_{ij}.$
One option is to directly route from $i$ to $j$ through the opaque network. If the shortest path from $i$ to $j$ is congested then it is possible to segment route  the connection from $i$ to some other edge node $k$ and then from $k$ to $j$. If the set of two shortest paths $i-k$ and $k-j$ are not congested, then this will result in better delay performance. 
Figure \ref{Fig: SR_S} shows an example of using segment routing to avoid a congested link. In this case, the link between router R1 and R2 is congested. We can route along the two 
segment path $R1-R3$, $R3-R2$ to avoid the congested link.
In general, we can route along a path with several segments. In this paper, we restrict the solutions to paths having at most two segments. The techniques developed extend directly to paths having more than two segments. Let $\beta_{ikj}$ represent the fraction of traffic from $i$ to $j$ that is routed though node $k$. The fraction of traffic that is routed on a single segment from  $i$ to $j$ is represented by $\beta_{ijj}$.   As in the egress picking case,
the delay suffered in the opaque network is a non-linear function of the traffic $T_{ij}$, the traffic splits $\beta_{ijk}$ as well as any other background traffic carried by the network. Let $\Delta_{ij}$ represent the delay on direct path between nodes $i$ and $j.$
We can think of this as the single hop segment delay. The delay $D_{ikj}$ incurred on the two segment path $i-k-j$ is $D_{ikj} = \Delta_{ik} + \Delta_{kj}.$ Assume that we have 
$n$ edge nodes in the opaque network.
We can then write the problem of finding the split values to minimize the 
average delay as 
\begin{equation}
    \min_{\bm{\beta}} \frac{\sum_{i=1}^{n} \sum_{j=1}^{m} \sum_{k=1}^{l} D_{ikj}\beta_{ikj} T_{ij}}{\sum_{i=1}^{n} \sum_{j=1}^{m} T_{ij}} 
\end{equation}
subject to
\begin{eqnarray*}
    \sum_{k=1}^{n} \beta_{ikj} & =  & 1, \forall i \; ,  \forall j \\
    \beta_{ikj} & \geq & 0 \forall i \; \forall j \; \forall k.
\end{eqnarray*}
The constraints that sets the sum of the traffic splits to one are the simplex constraints.
The main challenge in solving this problem is the fact that we do not know how the delay 
varies with the traffic split parameters. As in the egress picking problem, tomography 
provides end-to-end delay measurements that guides the machine learning based optimization algorithm.
We now outline the tomography based learning 
techniques that we use to solve this problem.
\section{Tomography Based Learning}
\label{methodology}
\begin{figure}[!t]
	\centering
	\includegraphics[width=.48\textwidth]{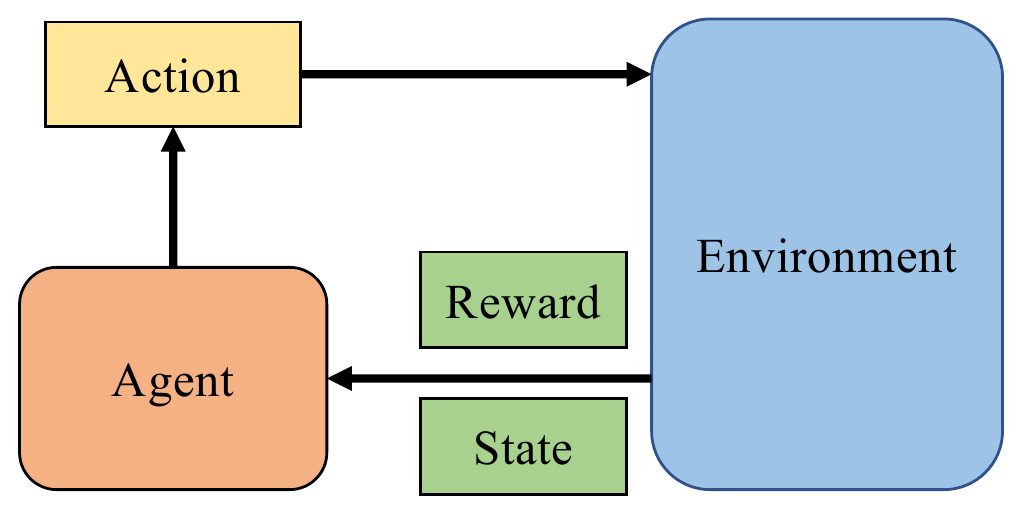}
\caption{A general setting for reinforcement learning}
\label{Fig: RL_S}
\end{figure}

The only information that we obtain from the opaque network is the tomography measurements. 
The idea is to use these measurements to determine how to distribute the load. The 
standard approach to using these measurements is in a learning based algorithm. 
The most straightforward approach is to use 
{\bf reinforcement learning (RL)}, where an agent interacts with an environment in discrete time steps. A general setting for RL is shown in Figure \ref{Fig: RL_S}. At each time step $t$, the agent observes the state of the environment $s_t$, takes certain action and receives reward $r_t$. The common objective for the agent is to maximize the expected cumulative discounted reward $E[\sum_{t=0}^{\infty} \gamma^t r_t]$.  In our case, the state is the current traffic split and the reward corresponds to the tomography inferred delays. The action that we take is to change the traffic split to optimize the objective function.
In order to convert the minimization problem to a maximization problem, the reward can be modeled as the negative of the weighed sum of mean delays.
RL maintains a table comprising of the received rewards for each state-action pair that 
has been used thus far. When a new state is observed, the action that results in the 
maximum reward is chosen in the exploitation mode or a random action is taken in the exploration mode.  For small and discrete state and action space, this can be done by using a simple table. However, for large and continuous state and action spaces, it is impossible to store the table directly. This is the case for our problems. 
Since deep neural networks have shown great potential in function approximation, it is possible to replace the state-action-reward table with a deep neural network. This technique, called {\bf Deep Q Network (DQN)}, has been used for solving problems with a large continuous state space \cite{mnih2015human}. In the DQN algorithm, a deep neural network is used to estimate the reward for each discrete action under a given state. At each step, the agent chooses the action with the maximum estimated reward. 
Though DQN shows great potential for solving problems with large state spaces, it can only solve problems with discrete and low-dimensional action spaces. In our case, the action space is the traffic split values which is continuous. For problems with continuous action space, DQN cannot be directly applied because the action is chosen based on discrete maximization of the estimated reward. Though the continuous action space can be discretized, the number of actions will increase exponentially with the number of degrees of freedom. For a continuous action space, the preferred approach is to use {\bf Deep Deterministic Policy Gradient (DDPG)} \cite{lillicrap2015continuous}.  DDPG comprises of two neural networks to 
determine the optimal action. An {\em actor network} determines the optimal action for 
a given state and a {\em critic network} estimates the reward for a given state-action pair. The actor is trained with the policy gradient provided by the critic network. Without special parameter tuning, DDPG has shown promising results on various continuous control problems \cite{lillicrap2015continuous}. However, DDPG is applicable when the action space is unconstrained and our problems have several constraints.

\subsection*{Enforcing Constraints}
In our case, the action space comprises of traffic splits and is constrained both by the 
simplex constraint (sum of the splits equals one) as well as non-negativity constraints. In addition, as stated earlier, there may be policy constraints that have to be enforced. Therefore, 
it will be convenient to use a method where it is easy to enforce constraints on the action space.
A technique that is suitable for these types of problems is the {\bf Critic Only Reinforcement Learning (CORL)} \cite{alibekov2018policy}. Unlike DDPG that uses two neural networks, the critic network for estimating rewards and an actor network for determining the optimal actions, CORL
trains only a critic neural network. The action or policy can be derived directly from the 
critic network by determining the action that minimizes the estimated cost provided by the 
critic. While solving the optimization problem we can enforce constraints on the solution
space. There are two ways to enforce the simplex constraints.
\begin{itemize}
    \item {\bf Enforcing Constraints using Softmax:} The standard approach to enforcing simplex 
    constraints in a neural network is to use the softmax function. The softmax function
    $f(x_i) = e^{x_i}/\sum_1^K e^{x_k}$ is enforced at the output layer of the 
    actor network. This ensures that sum of the probability over all the actions is one.
    However, the softmax function does not fully cover the entire action space. For example, the softmax function cannot set one of the outputs to one and all other outputs to zero.  However, the softmax function
    is simple to implement in a neural network and works reasonably well in practice to 
    enforce simplex constraints.
    \item{\bf Enforcing  Constraints using Projection or Frank-Wolfe:} An alternative approach is to start off with with a feasible operating point that satisfies all constraints (but may not be optimal). Then the constraints are explicitly enforced by either projecting the gradient or by the projection free Frank-Wolfe method to ensure that the 
    new operating point does not leave the feasible region. We show that by combining CORL with the Frank-Wolfe algorithm, we obtain rapid convergence to the optimal solution. To the best of our knowledge, this is the first use of Frank-Wolfe in CORL.
\end{itemize}

\subsection{Tomography and Load Distribution in DDPG and CORL}
For DDPG and CORL, the tomography module and load distributor correspond to different parts of the architecture. For both DDPG and CORL, the tomography module is the 
part of the critic network that measures the effect of a given action.
 The actor network serves as the load distributor in DDPG. 
 In CORL where there is no actor network, the critic network along with the constraint 
 enforcement module acts as the load distributor.
  During each interaction with the environment, the critic network estimates the best possible action, gets the corresponding reward from the environment using tomography, and learns from the reward-action pair. By directly serving as both the load distributor and the reward estimator, the critic network is able to probe the black box network more efficiently. According to our results, CORL converges faster and performs better than DDPG in most cases. 
\section{Critic Only Reinforcement Learning Methods for Traffic Splitting}
We describe the Critic Only Reinforcement Learning (CORL) algorithm as applied to our problem in more detail. First, we outline CORL where the simplex constraints are 
enforced using  a softmax layer. Since typical neural networks for classification tasks come equipped with 
the softmax function, this algorithm is easy to implement and performs reasonably well 
in terms of delay minimization on the topologies tested. Next we briefly outline the 
Frank-Wolfe algorithm that is used to enforce the constraints. The Frank-Wolfe based approach can be used to enforce linear constraints as long as we can solve a linear 
optimization problem over these constraints. In the case of the simplex and non-negativity constraints, the linear programming problem is trivial to solve, and this makes the 
Frank-Wolfe approach extremely attractive.
\subsection{Critic Only Reinforcement Learning}
The CORL algorithm has a critic neural network $Q(\bm{s}, \bm{a}| \theta)$, where $\bm{s}$ is the state, $\bm{a}$ is the action and $\theta$ is the parameter for the neural network. The critic network estimates the reward for a given pair of state and action. A suitable action given a state can be derived by performing gradient descent with the critic. The CORL method is described in Algorithm \ref{alg:CORL}. 
Note that for the problem of traffic splitting, the choice of action has no impact on the transition of states, since the traffic demands depend solely on the users. So instead of maximizing the discounted reward, at each time, only the current reward is maximized.
Standard CORL is used for maximization. Since our objective is to minimize mean delay, we 
maximize the negative of the mean delay.
\begin{algorithm}
\label{CORL}
\caption{CORL algorithm}
\label{alg:CORL}
\begin{algorithmic}[1]
\STATE Randomly initialize critic network $Q(\bm{s}, \bm{a}|\theta^Q)$ with weights $\theta^Q$. \\
\STATE Initialize target network $Q'$ with weights $\theta^{Q'} \leftarrow \theta^Q$. \\
\STATE Initialize replay buffer R. \\
\FOR{$t=0, \ldots, T$ } \do \\
\STATE Collect traffic demand $\bm{s}_t$ from the system. 
\STATE Generate one batch of random traffic split vectors $\bm{v}_0, ... \bm{v}_{N}$. \\
\STATE Enforce the constraint by setting $\bm{a}_i = f(\bm{v}_i)$
\STATE Set $\bm{v} = \argmin_{\bm{v}_i} Q(\bm{s_t}, f(\bm{v}_i) | \theta^{Q'}) $ \\
\FOR{$k=0, \ldots, K$ } \do \\
\STATE Update $\bm{v}$: $\bm{v} \leftarrow \bm{v} + \gamma \nabla_{\bm{v}} Q'(\bm{s}_t, f(\bm{v}) | \theta^{Q'})$. \\
\ENDFOR
\STATE Execute traffic split $\bm{a}=f(\bm{v})$. 
\STATE Collect information from the black box network and estimate the average delay $c_t$.
\STATE Store traffic demand, traffic split and average delay $(\bm{s}_t, \bm{a}_t, c_t)$ in R.
\STATE Sample a minibatch of $M$ buffer samples $(\bm{s}_i, \bm{a}_i, c_i)$ from R.
\STATE Update critic by minimizing the loss: $L = \frac{1}{M} \sum_i (c_i - Q(\bm{s}_i, \bm{a}_i | \theta^Q))^2$. In this case the critic is also the tomography module and it is capable of emulating the network. 
\STATE Update the target network: $\theta^{Q'} \leftarrow \tau \theta^Q + (1-\tau)\theta^{Q'}$.\\
\ENDFOR
\end{algorithmic}
\end{algorithm}
Similar to DDPG \cite{lillicrap2015continuous}, we use the "soft" update mechanism for the critic network. A copy of the critic network is created as the target critic network $Q(\bm{s}, \bm{a}| \theta)$. The weights of the target network is updated slowly according to the learned critic network: $\theta^{Q'} \leftarrow \tau \theta^Q + (1-\tau)\theta^{Q'}$ with $\tau \ll 1$. At the beginning of the experiment, both networks are initialized with the same random weights. A fixed length replay buffer is used to store past action, state and reward data for training. At the beginning of each time slot, a batch of random vectors that correspond to traffic splits $\bm{v}_0, ... \bm{v}_N$ is generated. The softmax activation function $f(x_i) = e^{x_i}/\sum_1^K e^{x_k}$ is used on the corresponding elements of $\bm{v}$ to enforce the simplex and non-negativity constraints. An initial split is selected by picking the vector with the lowest estimated cost. Then the action is optimized for $K$ iterations using the gradient provided by the target critic network. Finally the traffic split is executed and a reward is estimated from the environment. After each interaction with the environment, the original demand, traffic split and average delay is stored in the buffer. A minibatch of data is selected from the buffer to update the critic target by minimizing the MSE of the estimated average delay. At the end of each time slot, the target critic network is "soft" updated. 
Note that for the problems we consider in this paper, the states are represented by a traffic demand matrix. Since the action taken by the agent has no impact on the traffic demand, so the agent is trying to minimize the average delay at current time. 
We now outline the Frank-Wolfe algorithm in general and then show how we incorporate it 
into CORL.
\subsection{The Frank-Wolfe Algorithm}
The Frank-Wolfe algorithm \cite{frank1956algorithm, jaggi2013revisiting} was proposed to solve convex optimization problems over linear polytopes. The optimization problem that 
we want to solve is 
\begin{equation}
    \min_{\bm{x} \in \mathcal{D}}  f(\bm{x}),
\end{equation}
where $\mathcal{D}$ is a linear polytope. The Frank-Wolfe algorithm starts off at an initial feasible point. If the polytope $\mathcal{D}$ is complicated, then finding a feasible point itself is non-trivial. For our problem, any arbitrary set of traffic splits
is feasible. The algorithm iterates through a sequence of feasible points approaching the optimal solution.  At each step of the algorithm, the non-linear objective function is linearized at the current feasible point. Next, a linear programming problem is solved 
with this linear objective function over the polytope $\mathcal{D}.$ This solution 
will be an extreme point of $\mathcal{D}.$ We then move along the straight line from the current feasible point to the current optimal extreme point. We can either perform a line 
search to determine the optimal point to move to or we can use a step length function that
guarantees convergence. We use the second approach. A description of the algorithm is 
given in Algorithm \ref{alg:FW}.
\begin{algorithm}
\label{FW}
\caption{Frank-Wolfe}
\label{alg:FW}
\begin{algorithmic}[1]
\STATE Pick an arbitrary $\bm{x_0} \in \bm{D}$ \\
\FOR{$k=0, \ldots, K$ } \do \\
\STATE Compute $\bm{z} := \argmin_{\bm{t} \in \bm{D}} \langle \bm{z}, \nabla f(\bm{x}_k)  \rangle$ \\
\STATE $\bm{x}_{k+1} = (1-\gamma)\bm{x}_k + \gamma \bm{z}, \gamma = \frac{2}{k+2}$  \\
\ENDFOR
\end{algorithmic}
\end{algorithm}
The solution at step $k$ satisfies $f(\bm{x}_k) - f(\bm{x}^*) \leq \mathcal{O}(\frac{1}{k})$, where $\bm{x}^*$ is the optimal solution. 
In \cite{mokhtari2018stochastic}, the authors show that even with noisy estimates of the gradient, the Frank-Wolfe based method can achieve a bounded approximation of the optimal solution for several types of linear polytopes. We now show how to incorporate the
Frank-Wolfe algorithm into CORL. We call this algorithm CORL-FW.
\subsection{Critic Only Reinforcement Learning with Frank-Wolfe Optimization}
We now outline CORL-FW, which combines Frank-Wolfe with CORL. 
In the standard Frank-Wolfe algorithm, the gradient of the non-linear objective function
is computed at the current operating point. Since the network is opaque, we do not know
the objective function. Therefore, we use the 
critic network to provide the estimate of the gradient for the Frank-Wolfe method.
The linear programming problem is trivial to solve for both the representative problems. 
In the case of egress picking, the linear programming problem is separable over the 
different prefixes and in the traffic engineering problem the linear program is separable over different traffic source-destination pairs. Once the optimal solution is determined, 
the new traffic splits are computed by using the step length shown in Algorithm  \ref{alg:FW}.
Experimental results show that even with estimated gradients CORL-FW can achieve close to optimal solutions. 
Details of CORL-FW is shown in Algorithm \ref{alg:CORL-FW}. 
\begin{algorithm}
\label{CORL-FW}
\caption{CORL-FW algorithm}
\label{alg:CORL-FW}
\begin{algorithmic}[1]
\STATE Randomly initialize critic network $Q(\bm{s}, \bm{a}|\theta^Q)$ with weights $\theta^Q$. \\
\STATE Initialize target network $Q'$ with weights $\theta^{Q'} \leftarrow \theta^Q$. \\
\STATE Initialize replay buffer R. \\
\FOR{$t=0, \ldots, T$ } \do \\
\STATE Collect traffic demand $\bm{s}_t$ from the system. 
\STATE Generate one batch of random traffic splits $\bm{a}_0, ... \bm{a}_{N}$ from the action space $ \mathcal{D}$. 
\STATE Set $\bm{a} = \argmin_{\bm{a}_i} Q(s, \bm{a}_i | \theta^{Q'}) $ \\
\FOR{$k=0, \ldots, K$ } \do \\
\STATE Compute $\bm{z}: = \argmin_{\bm{z} \in  \mathcal{D}} \langle \bm{z} , \nabla_{\bm{a}} Q'(\bm{s}, \bm{a} | \theta^{Q'})$
\STATE $\bm{a} \leftarrow (1-\gamma)\bm{a} + \gamma \bm{z}, \gamma = \frac{2}{k+2}$
\ENDFOR
\STATE Execute traffic split $\bm{a}$.  
\STATE Collect information from the black box network and estimate the average delay $c_t$.
\STATE Store original demand, traffic split and average delay $(\bm{s}_t, \bm{a}_t, c_t)$ in R.
\STATE Sample a minibatch of $M$ buffer samples $(\bm{s}_i, \bm{a}_i, c_i)$ from R.
\STATE Update critic/tomography module by minimizing the loss: $L = \frac{1}{M} \sum_i (c_i - Q(\bm{s}_i, \bm{a}_i | \theta^Q))^2$. 
\STATE Update the target network: $\theta^{Q'} \leftarrow \tau \theta^Q + (1-\tau)\theta^{Q'}$.\\
\ENDFOR
\end{algorithmic}
\end{algorithm}

\section{Experimental Results}
We test the performance of our methods on real typologies from the Rocketfuel project \cite{spring2004measuring} and Abilene dataset \cite{zhang2003fast}. Note that this explicit knowledge of topologies is used only for evaluation purposes, i.e., to have the ground truth for evaluation purposes. Clearly, the traffic sources do not use this information in their load distribution decisions, since from their perspective, the network to which they are sending traffic is a black box, and the only usable information about the network is that inferred from external probing (network tomography). The delay between source node $i$ and destination node $j$, $D_{i,j}$, consists of queuing delay and propagation delay on each link along the path. We use a common non-linear model from queueing analysis 
\begin{equation} \label{pdfeq1}
  g(x)  =
  \begin{cases}
   \frac{w}{1-x/C}+p & \text{ If } x < C \\
    D+p & \text{ If } x \geq C  
  \end{cases}
\end{equation}
to model the delay on a given link. Here $w= 1/\mu$ and $\mu$ is the service rate, $p$ is the fixed propagation delay and $D$ is a fixed congestion delay if the utilization of the link gets close to or greater than its capacity. In our experiments, we set $D$ to one second, so no link will have a queuing delay greater than one second. Again, it is worthwhile to stress that explicit knowledge of these parameters is used only for evaluation purposes. This knowledge is not used for determining optimal load distribution. 

To model realistic network traversal, for all the experiments, we assume ECMP is used throughout the black box network. So the overall end to end delay is a weighted sum of the delay from all the paths between the source and destination node. 

For both sets of topologies, we evaluate the effectiveness of our load distribution scheme for the two representative use cases (of routing through black box networks) discussed in detail earlier. For egress picking problems, all the neural networks for DDPG, CORL and CORL-FW are simple fully connected networks, with two hidden layers of size 256. For the case of segment routing, all the neural networks have two hidden layers of size 512 and 256. The size of the first layer is increased to cover the wider range of delays caused by congestion. 

For the DRL methods, to select a suitable initial point, 1000 random initial points are first evaluated with the critic networks, then the best one is selected as the initial point. All the DRL agents maintain a replay buffer with 1000 most recent states, actions and rewards. After each interaction with the environment, a batch of 32 samples are used for the update of the neural networks. A learning rate of 0.001 is used for all the neural networks. For CORL, we use the Adam optimizer for optimization of the action with a learning rate of 0.05. 

For CORL, at each time the optimization is run for 100 iterations. For CORL-FW, we set the stopping criteria to be either reaching 10 steps of optimization or when the Euclidean distance between $\bm{D}$ and $\bm{x}_k$ is under 0.00001. To derive a close lower bound, for the Frank-Wolfe (FW) method we assume that optimization is performed with accurate gradients for 100 iterations or until the distance between $\bm{D}$ and $\bm{x}_k$ is under 0.00001.

\subsection{ Results of Experiments with Rocketfuel Topologies}
We first show results comparing the performance of the different load distribution methods when each of five Rocketfuel topologies is used as the topology of the black box network. Details of the five topologies are shown in Table \ref{tab:topogolies}. 

\begin{table}
\renewcommand{\arraystretch}{1.0}
\caption{Rocketfuel Topologies}
\label{tab:topogolies}
\begin{center}
\begin{tabular}{ c || c c c c c }
\hline
Topology & rf1221 & rf1755 & rf3257 & rf3967 & rf6461 \\
\hline
Number of Nodes & 104 & 87 & 161 & 79 & 138 \\
\hline
Number of Links & 302 & 322 & 656 & 294 & 744 \\
\hline
\end{tabular}
\end{center}
\end{table}

Since real traffic matrices (TMs) are not available for these five topologies, we randomly generate TMs using the gravity model \cite{roughan2005simplifying}. The gravity model assumes demand $p_{ij}$ from node $i$ to node $j$ is, 
\begin{equation}
p_{ij} = p^{in}_i p^{out}_j
\end{equation}
where $p^{in}_i$ and $p^{out}_j$ can be randomly generated from an exponential distribution for each node. To model the correlation across time, we assume that for the duration of the experiment, at each time each $p^{in}_i$ and $p^{out}_j$ is drawn from a Guassian distribution. In our experiments, we first generate the mean values for each $p^{in}_i$ and $p^{out}_j$, and scale up the mean values so that the maximum link utilization is over 90 percent (to effectively illustrate the delay impact). We assume that for the duration of the experiment, the TMs are relatively stable, so that the standard deviations are within one percent of the mean values.

\begin{figure*}[ht]
\subcaptionbox{rf1221}{%
\includegraphics[width=0.208\textwidth]{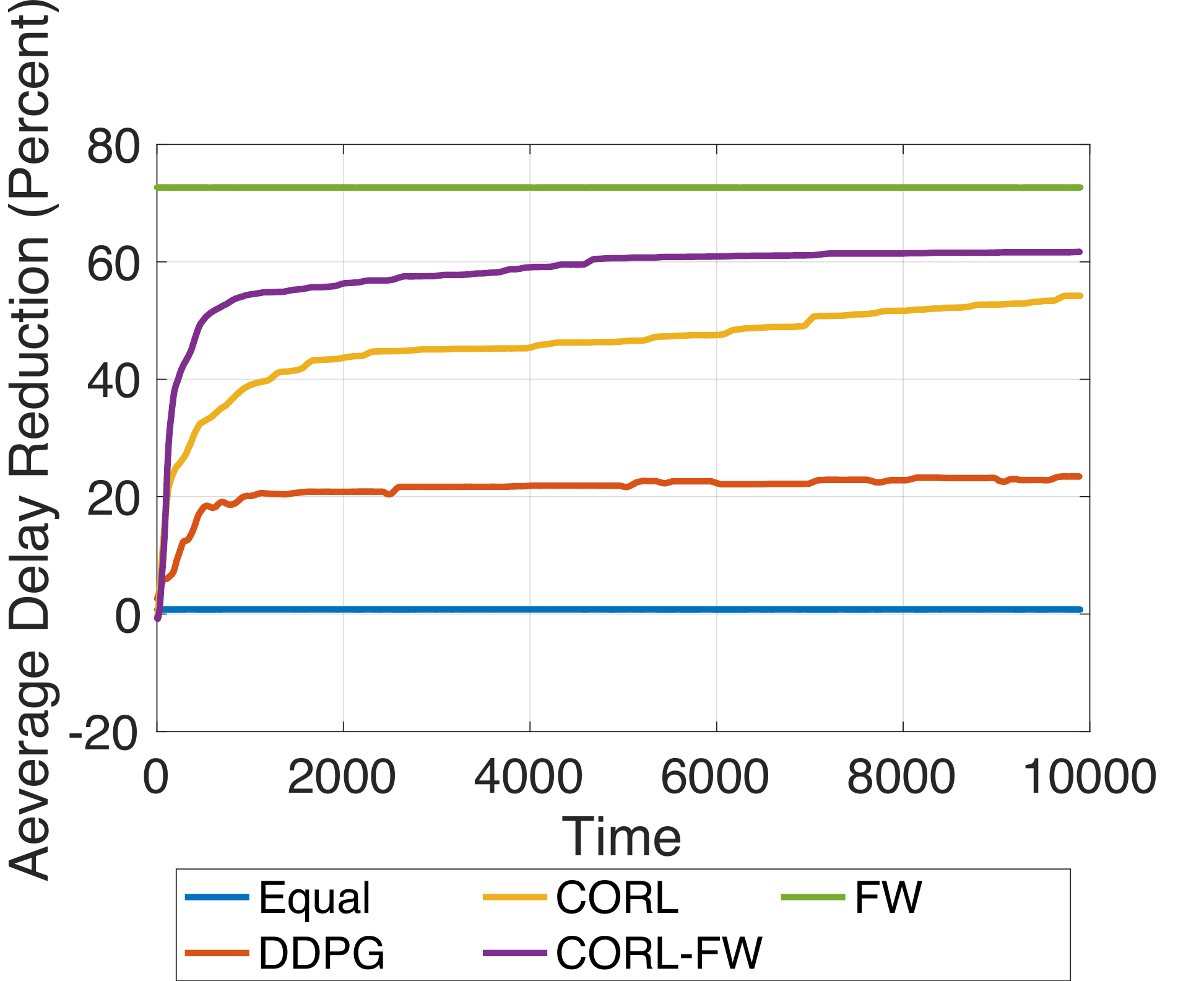}}%
\hfill
\subcaptionbox{rf1755}{%
\includegraphics[width=0.198\textwidth]{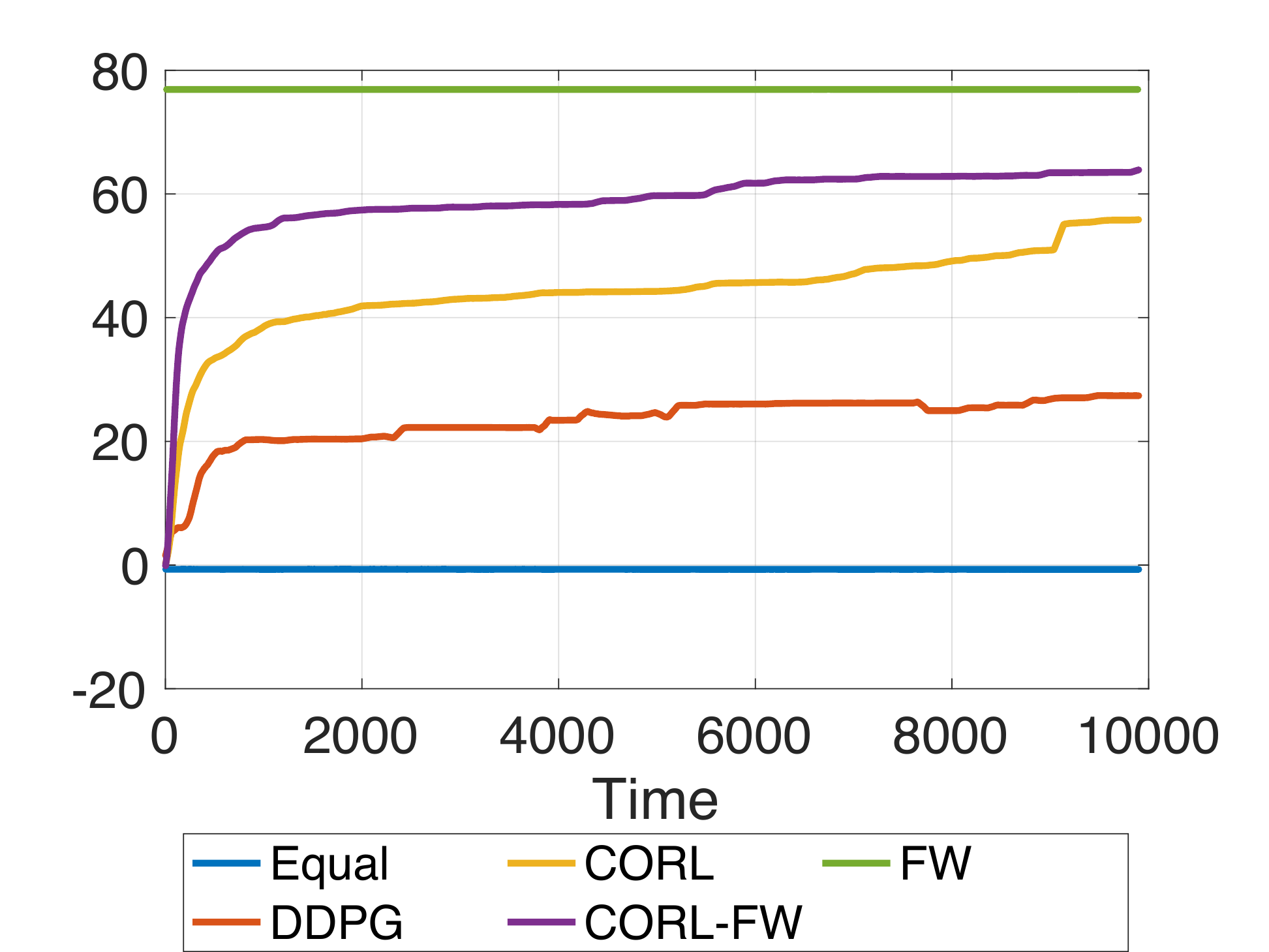}}%
\hfill
\subcaptionbox{rf3257}{%
\includegraphics[width=0.198\textwidth]{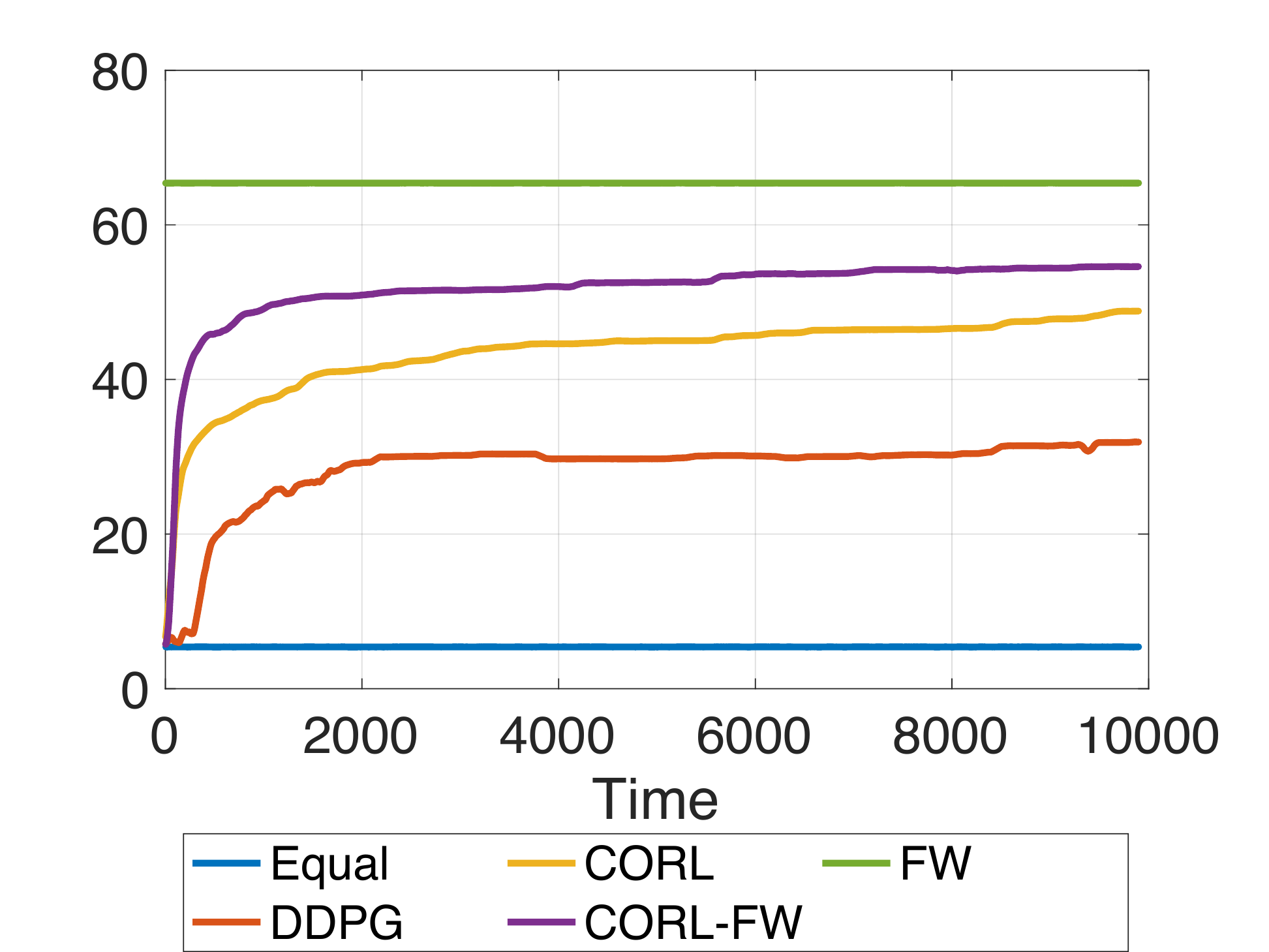}}%
\hfill
\subcaptionbox{rf3967}{%
\includegraphics[width=0.198\textwidth]{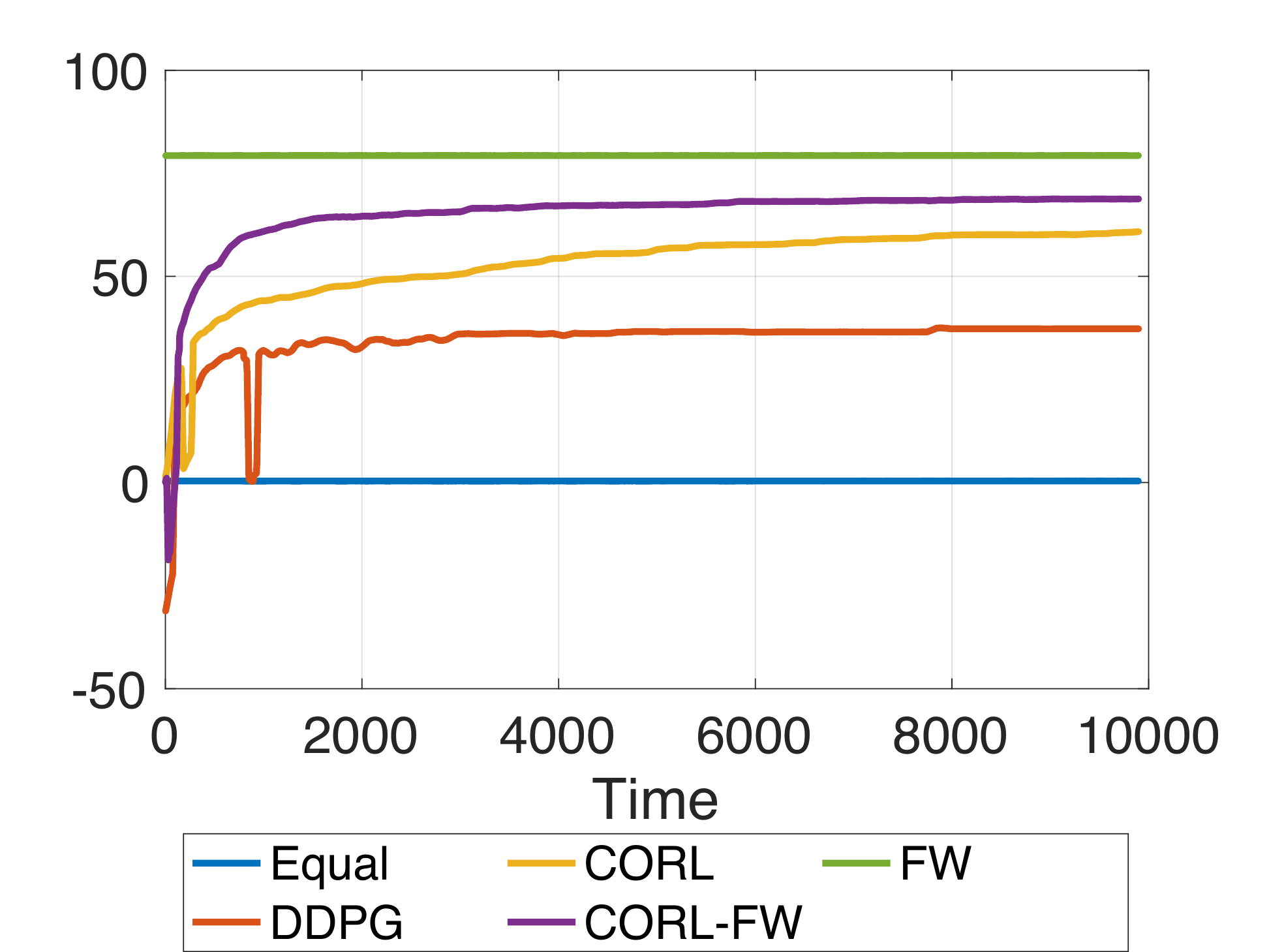}}%
\hfill
\subcaptionbox{rf6461}{%
\includegraphics[width=0.198\textwidth]{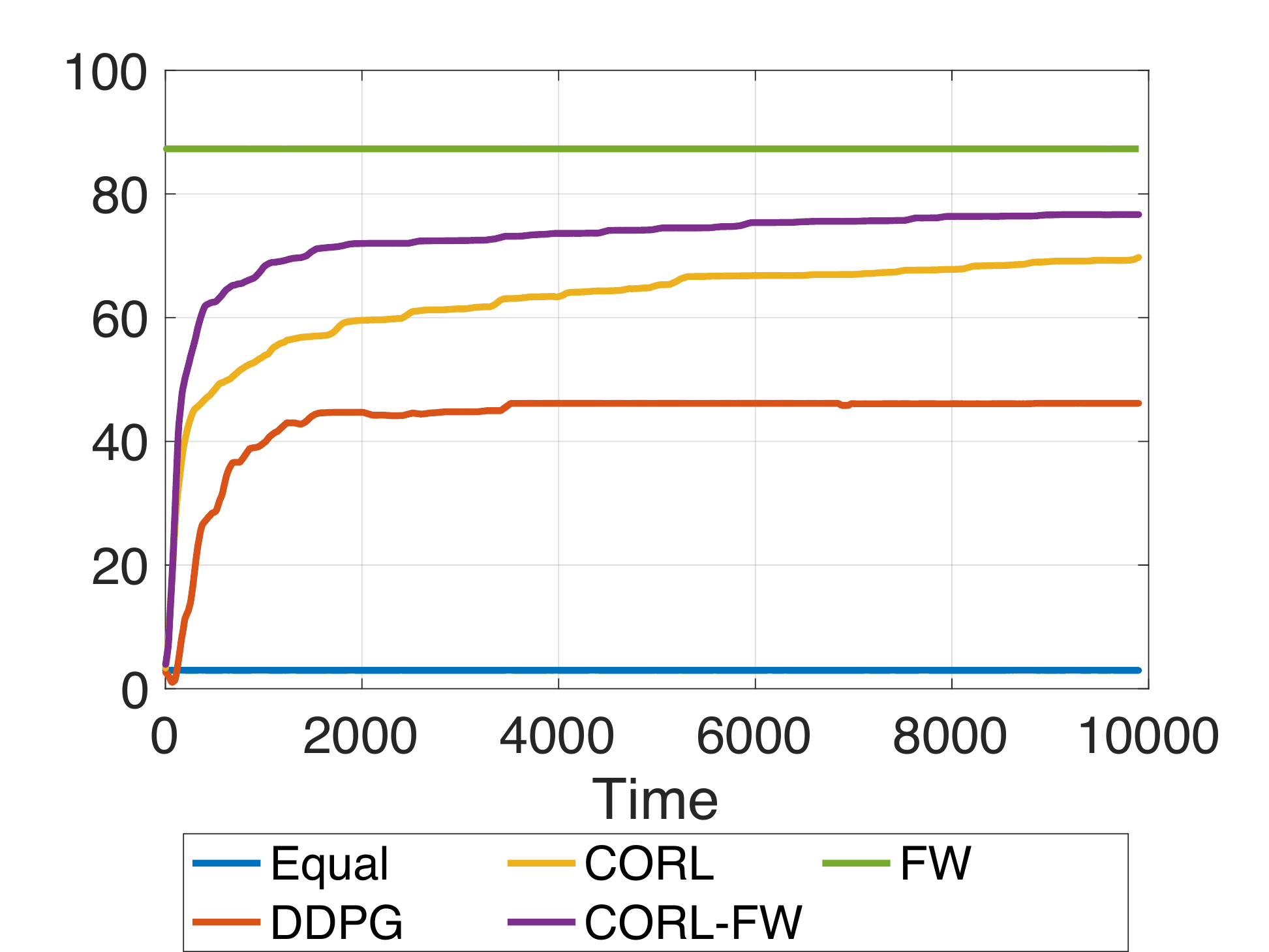}}%
\caption{Egress Picking}
\label{Fig:rf_ep}
\end{figure*}

For egress picking, we run ten experiments on each topology. For each run, 20 egress points are randomly selected; these are egresses from the sending ISP and implicitly correspond to the ingresses into the downstream black box network. 20 other nodes in the black box network are randomly selected as destinations -- these correspond to to true destinations or egresses from the black box network toward the final destinations. Again note that knowledge of these exit nodes from the black box network is for evaluation purposes only. Egress picking is performed by the sending ISP only for 20 egresses from its network. The black box network also has background traffic which we generated from the randomly chosen traffic matrix. Figure \ref{Fig:rf_ep} shows, for each topology and time instant, the reduction in average delay through the black box network (between its ingress nodes and destination nodes) for ten runs. 

For all the five topologies, CORL-FW achieves around 60 percent average delay reduction, with a gap of less than 20 percent from the lower bound. CORL-FW also achieves the highest performance improvement in the least amount of time. CORL achieves around 50 percent average delay reduction. While DDPG also reduces the average delay, it performs worse than the CORL methods. The naive heuristic of equally splitting traffic amongst the egress points hardly improves and sometimes even degrades performance. The results clearly show the effectiveness of using network tomography combined with learning-based decision making. Despite the network being a black box, considerable performance improvement is possible by judicious load distribution using only externally observable information.

\begin{figure*}[ht]
\subcaptionbox{rf1221}{%
\includegraphics[width=0.208\textwidth]{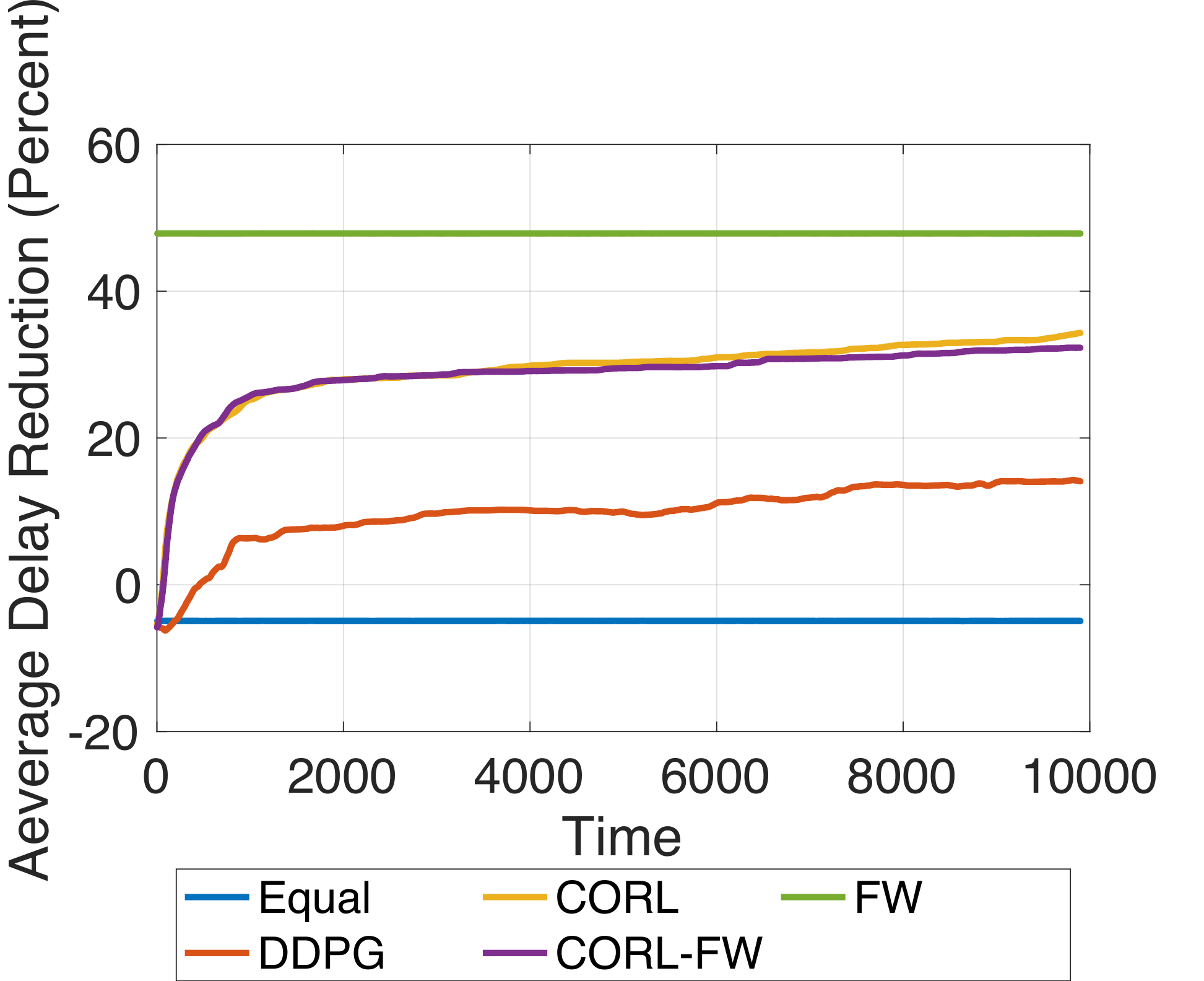}}%
\hfill
\subcaptionbox{rf1755}{%
\includegraphics[width=0.198\textwidth]{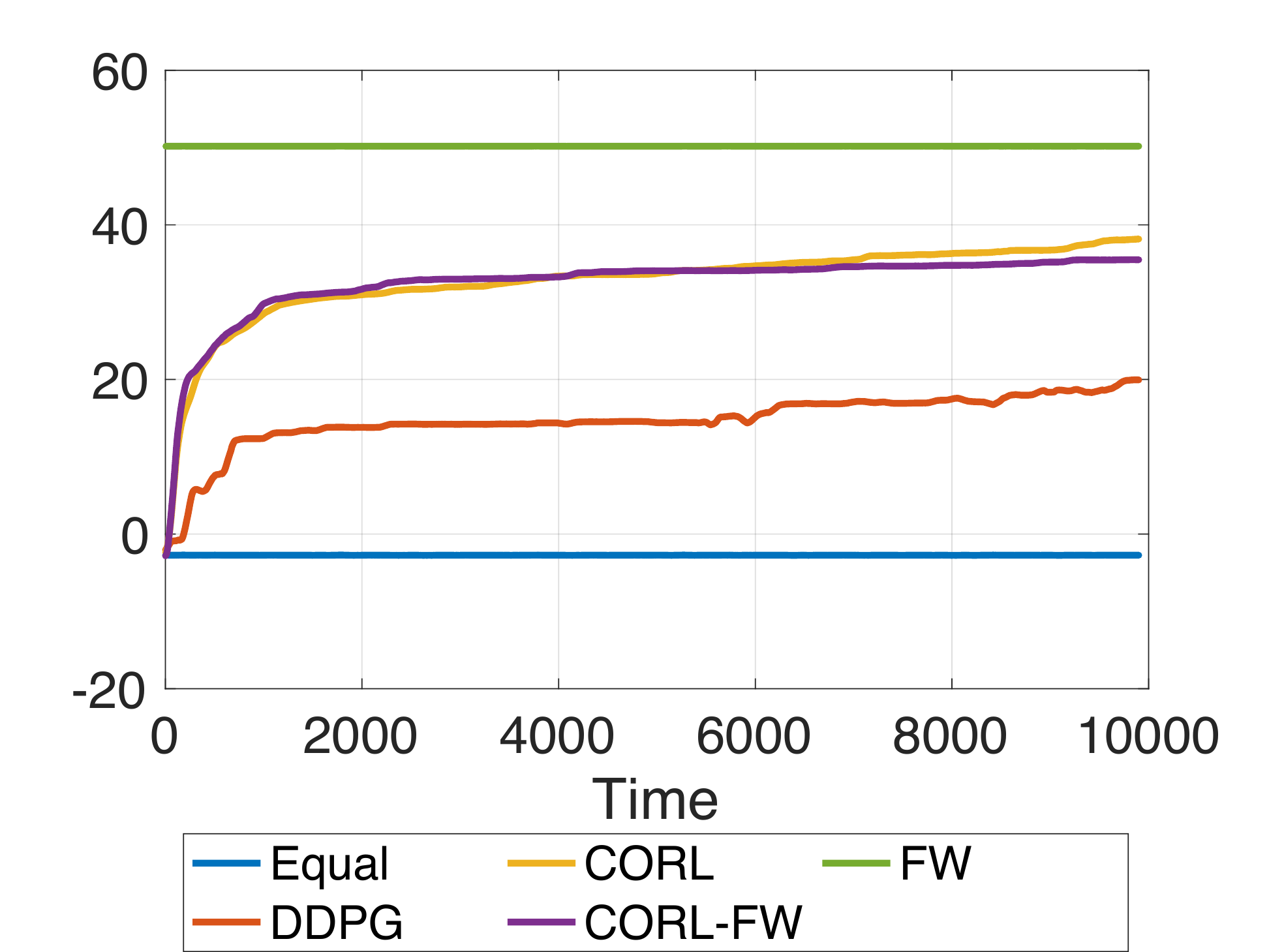}}%
\hfill
\subcaptionbox{rf3257}{%
\includegraphics[width=0.198\textwidth]{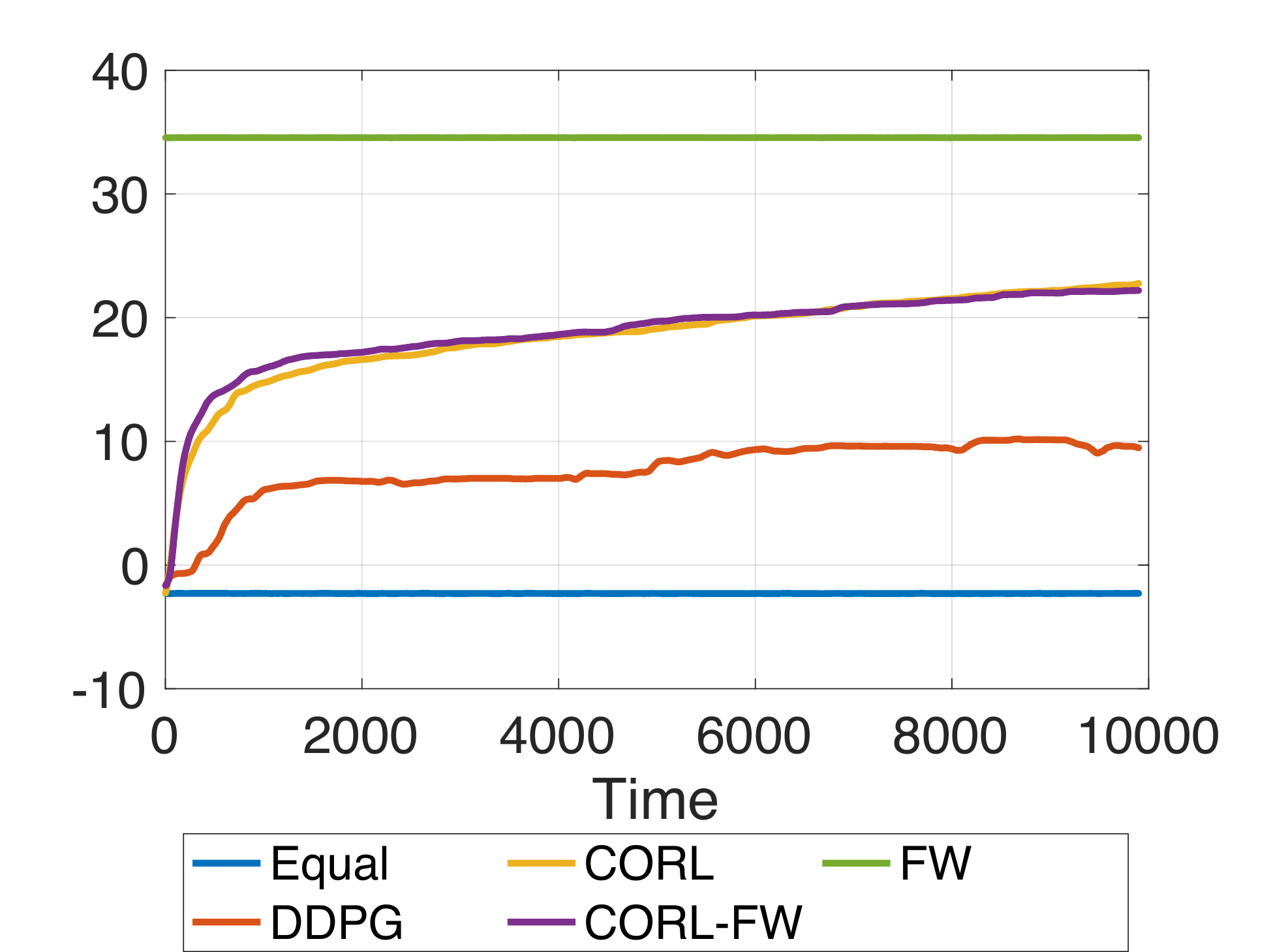}}%
\hfill
\subcaptionbox{rf3967}{%
\includegraphics[width=0.198\textwidth]{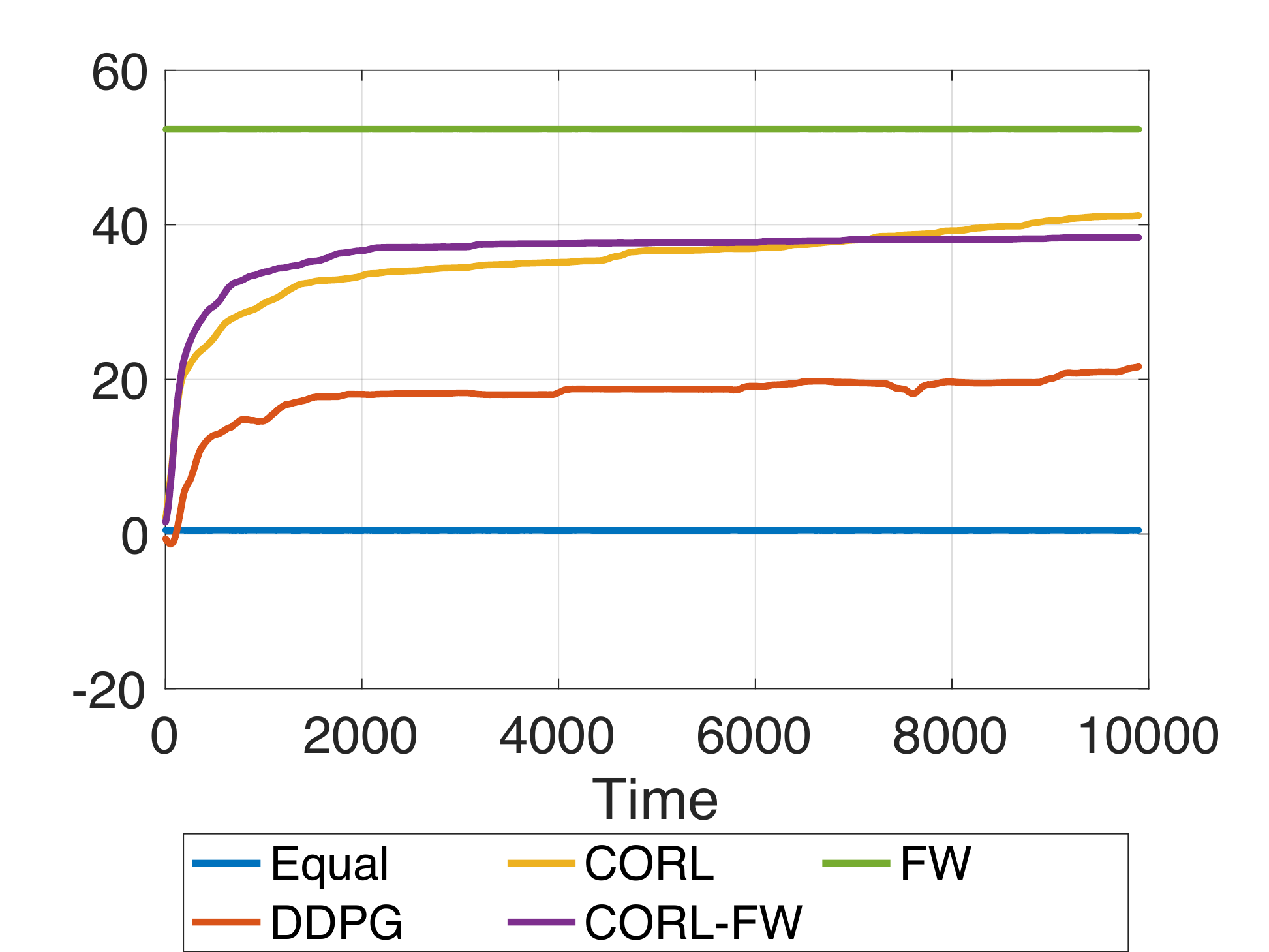}}%
\hfill
\subcaptionbox{rf6461}{%
\includegraphics[width=0.198\textwidth]{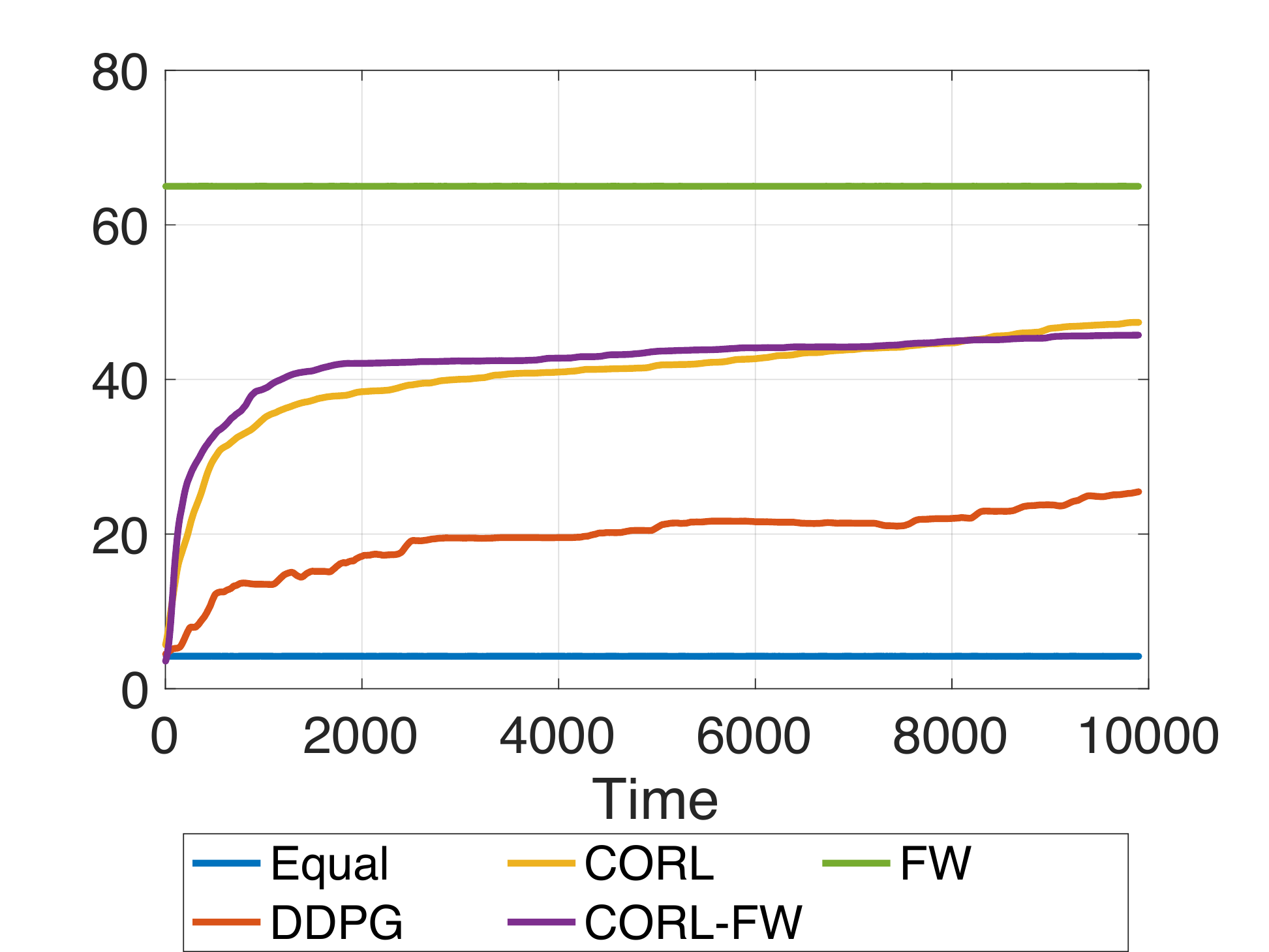}}%
\caption{Egress Picking with Four Egress Points}
\label{Fig:rf_ep_ao}
\end{figure*}

To evaluate the performance of the methods on large scale problems, we run another experiment in which four egress points are randomly selected, and all the other nodes are chosen as destinations. In this case there are 100, 83, 157, 75 and 134 prefixes for topology rf1221, rf1755, rf3257, rf3967 and rf6461, respectively. Results are shown in Figure \ref{Fig:rf_ep_ao}. With less egress points to choose from and more prefixes, the performance gains are lower than the previous case. However the CORL methods still perform consistently better than DDPG. With larger problem sizes, the performance of CORL is very close to CORL-FW. This may be because the hidden layers of the NNs is kept at 256, and the capacity of the NNs is bounding the performance of CORL-FW. 

\begin{figure*}[ht]
\subcaptionbox{rf1221}{%
\includegraphics[width=0.208\textwidth]{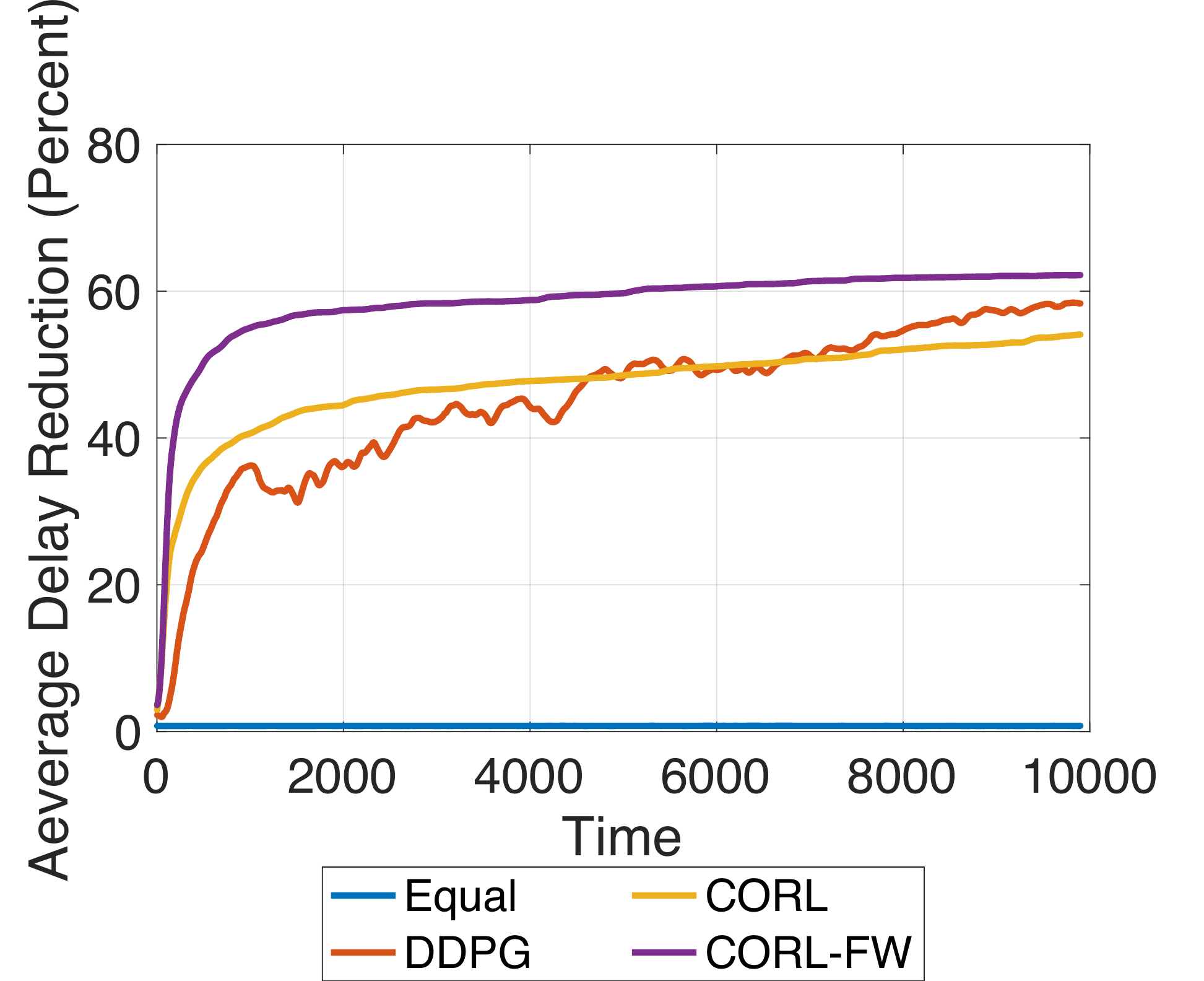}}%
\hfill
\subcaptionbox{rf1755}{%
\includegraphics[width=0.198\textwidth]{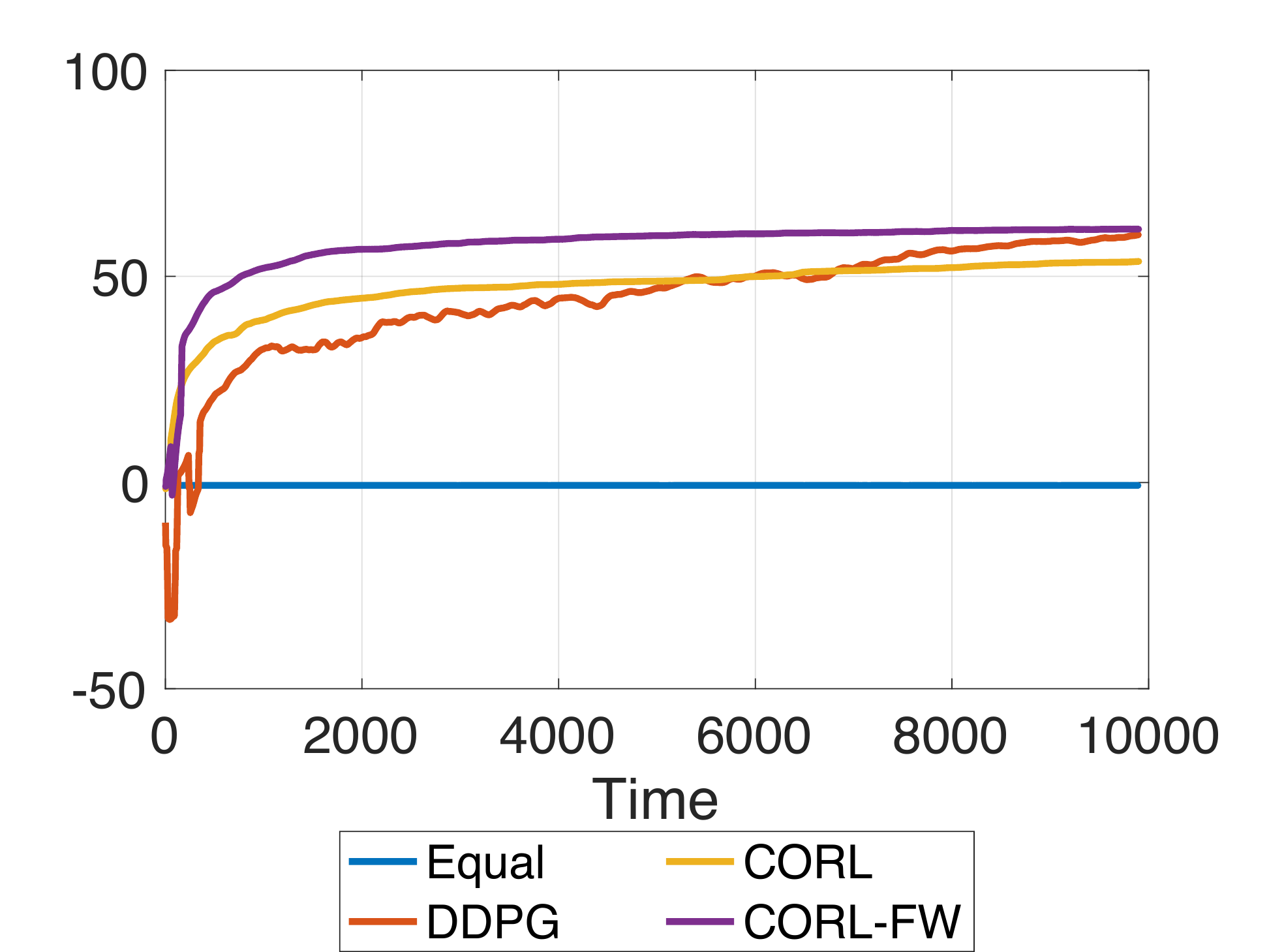}}%
\hfill
\subcaptionbox{rf3257}{%
\includegraphics[width=0.198\textwidth]{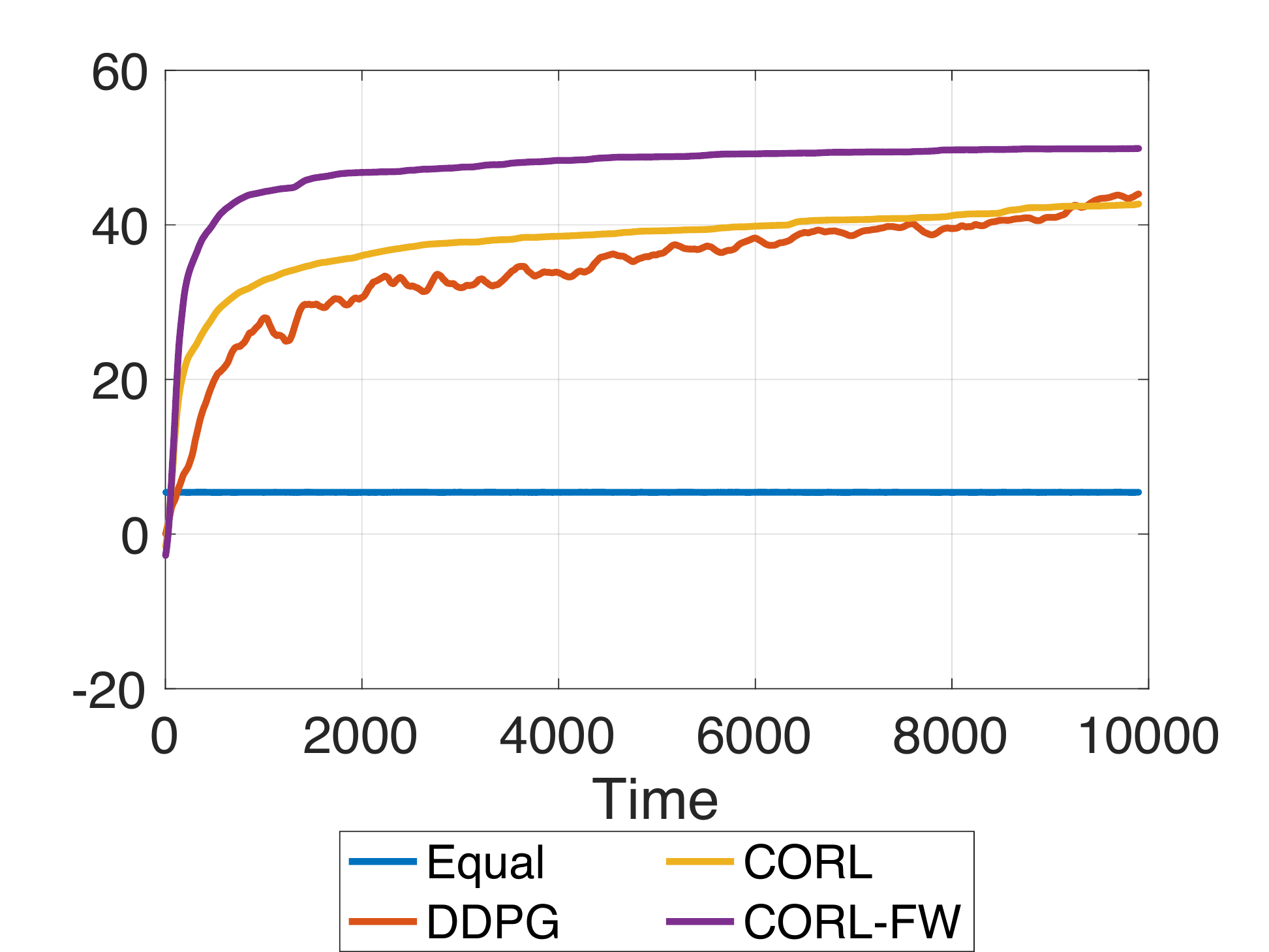}}%
\hfill
\subcaptionbox{rf3967}{%
\includegraphics[width=0.198\textwidth]{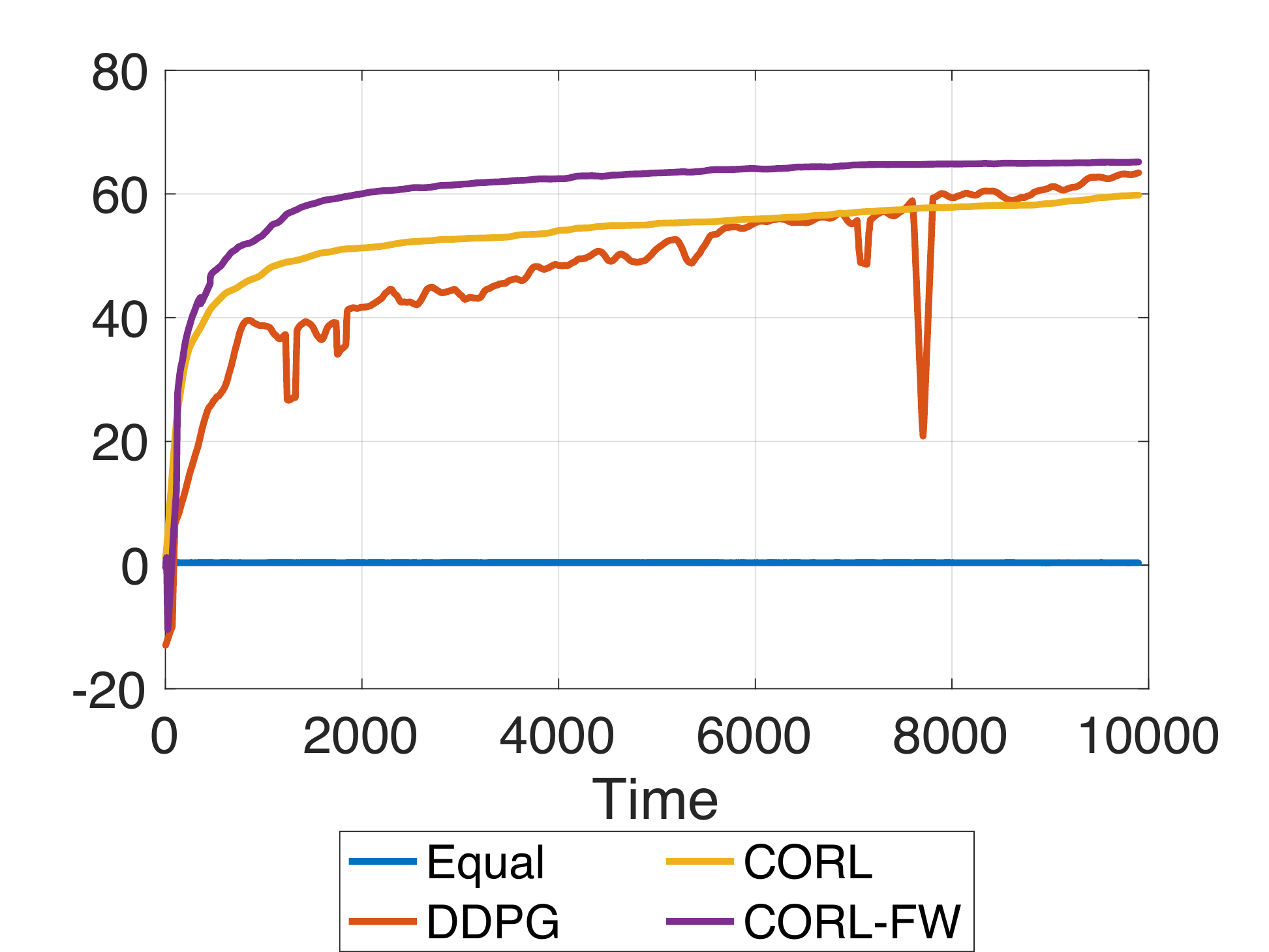}}%
\hfill
\subcaptionbox{rf6461}{%
\includegraphics[width=0.198\textwidth]{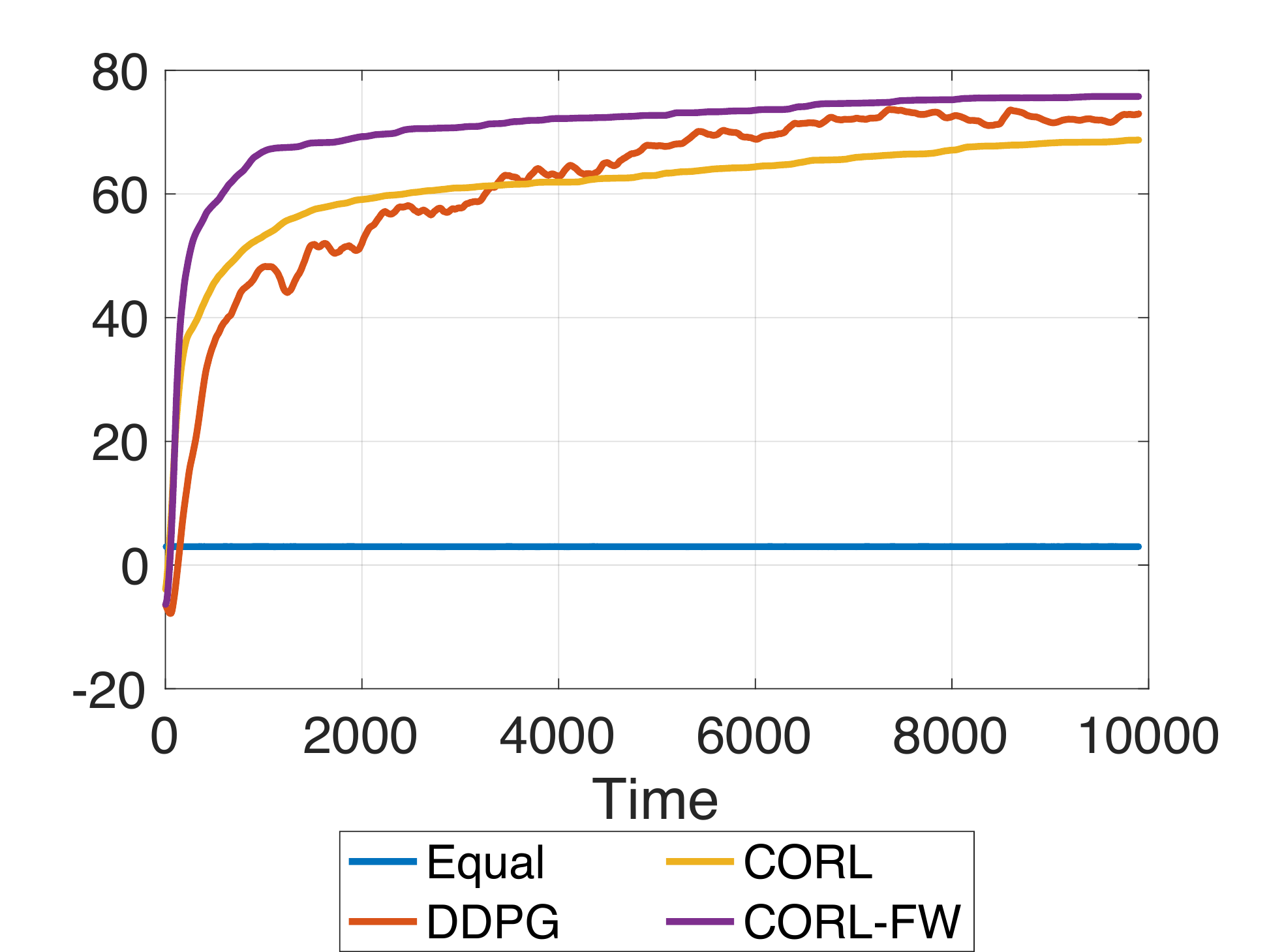}}%
\caption{Distributed Egress Picking}
\label{Fig:rf_EP_d}
\end{figure*}

Next, we test the performance when different sending ISPs simultaneously send traffic (using independent egress picking on their networks) to a common black box network. 
We assume four service providers are sending traffic using optimized egress picking. Each ISP picks 20 egresses for traffic toward 20 prefixes. One exception is for rf3967 where each service provider uses 19 egress nodes and 19 prefixes. Results are shown in Figure \ref{Fig:rf_EP_d}. 

For this distributed egress picking, CORL-FW outperforms all other methods in terms of average delay reduction and efficiency. For DDPG the overall performance actually oscillates with time -- this may be caused by the instability of the agents. The DDPG agents possibly were not able to learn the suitable policies to cooperate with others and were competing over the same link resources. This again shows the gains that can be achieved by combining the limited tomography derived information with learning-based decision making, even in a distributed setting with independent decision makers, as would be the case when multiple autonomous systems independently optimize their egress picking to a common opaque network.

\begin{figure*}[ht]
\subcaptionbox{rf1221}{%
\includegraphics[width=0.208\textwidth]{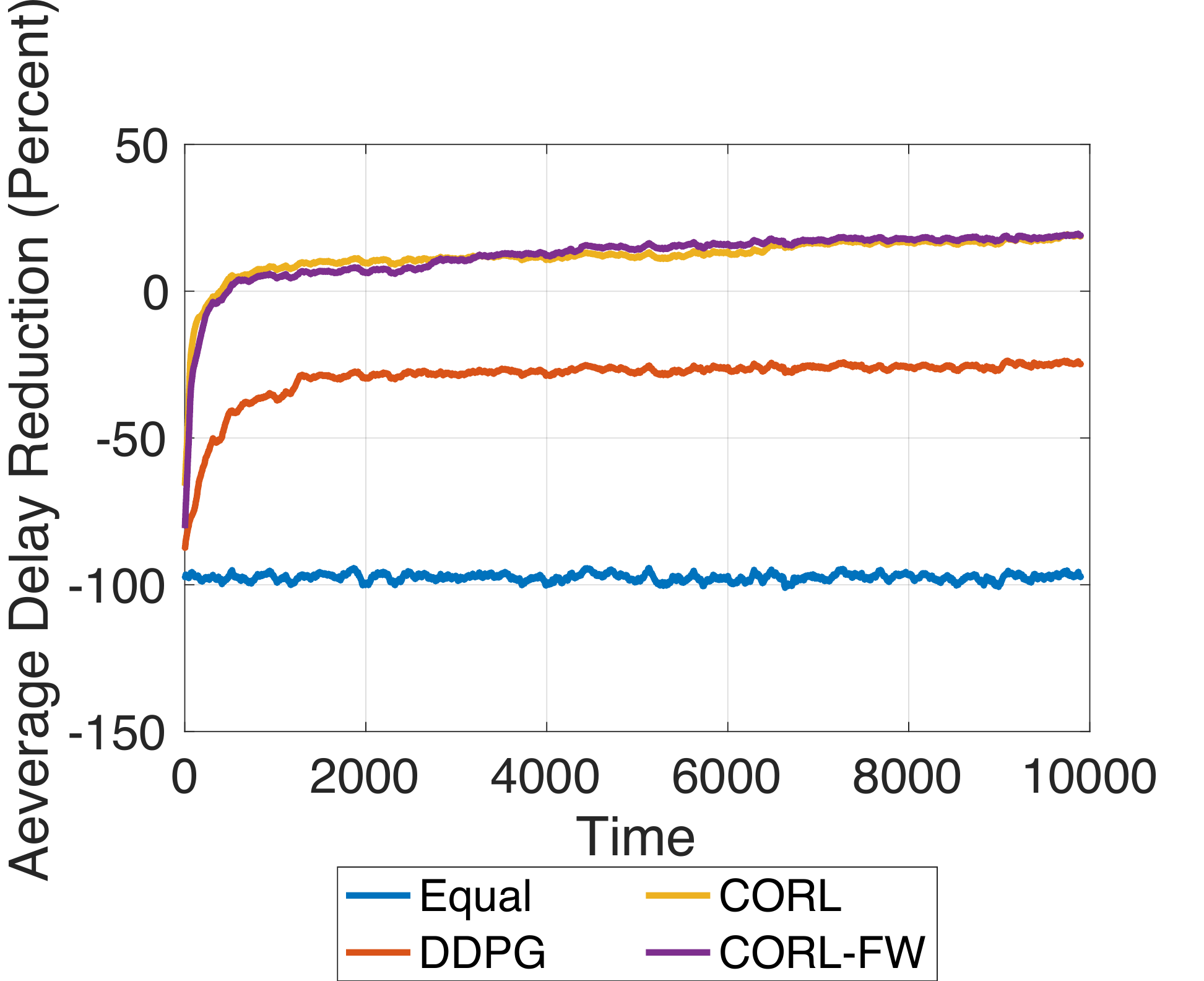}}%
\hfill
\subcaptionbox{rf1755}{%
\includegraphics[width=0.198\textwidth]{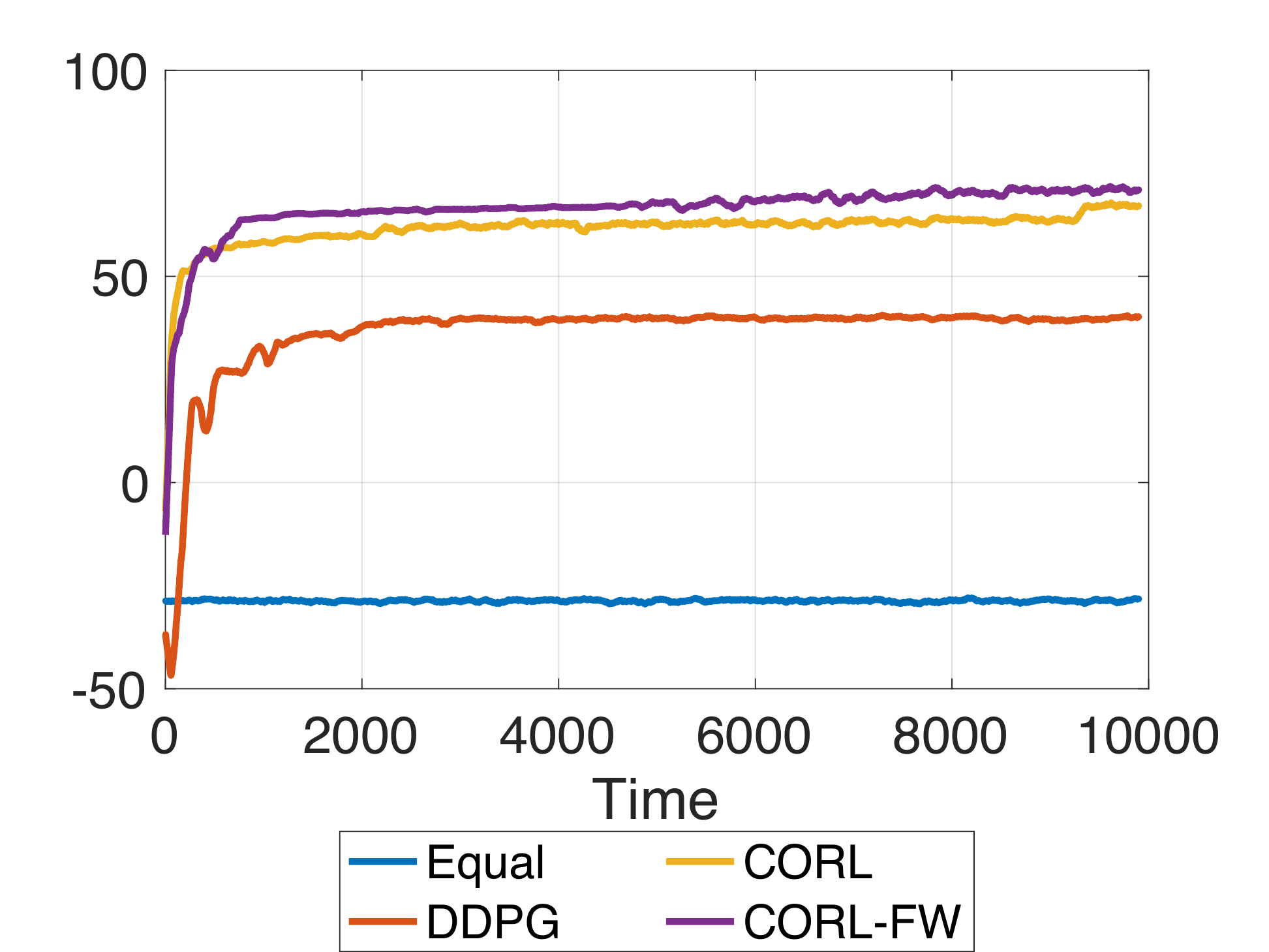}}%
\hfill
\subcaptionbox{rf3257}{%
\includegraphics[width=0.198\textwidth]{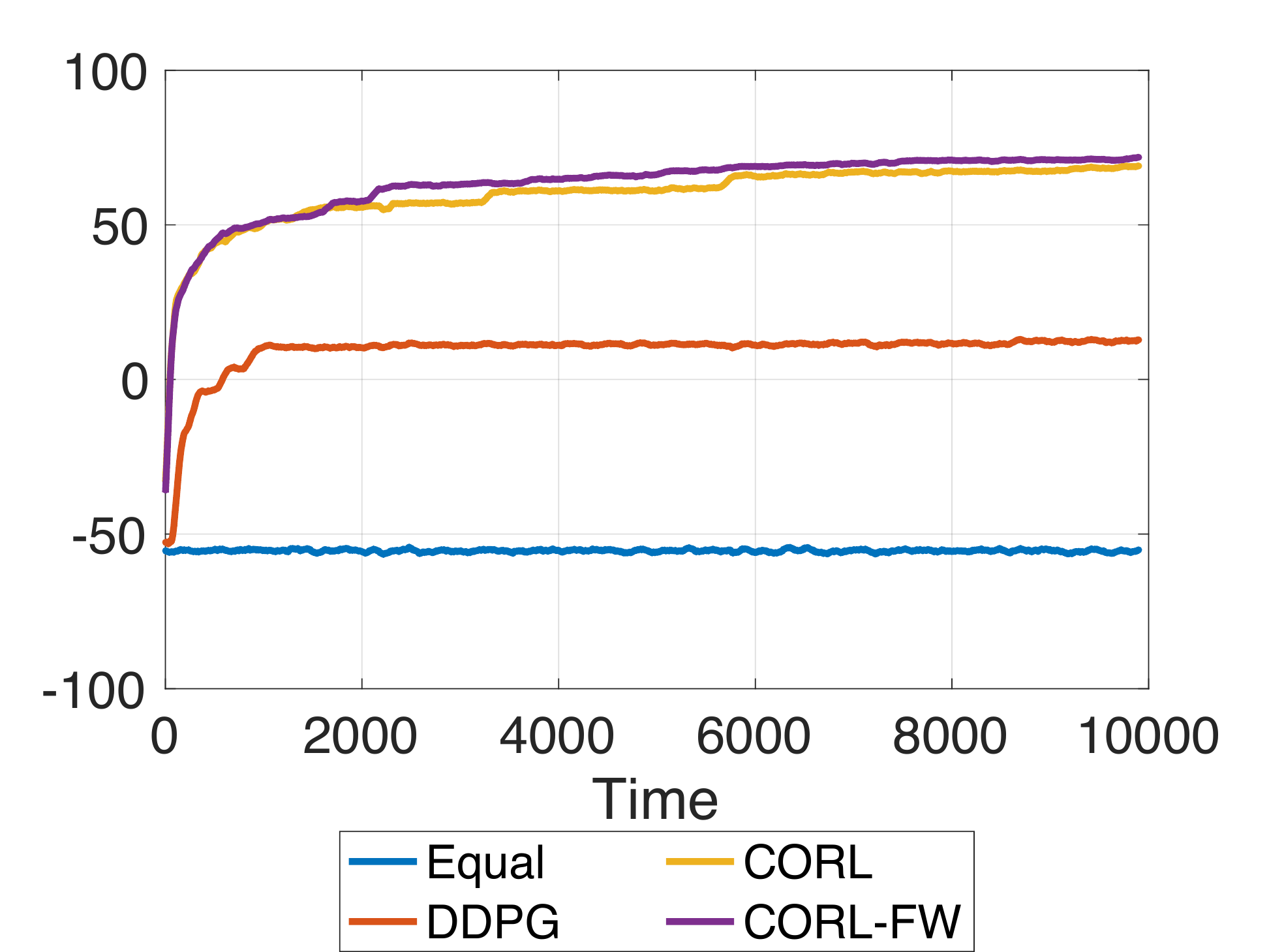}}%
\hfill
\subcaptionbox{rf3967}{%
\includegraphics[width=0.198\textwidth]{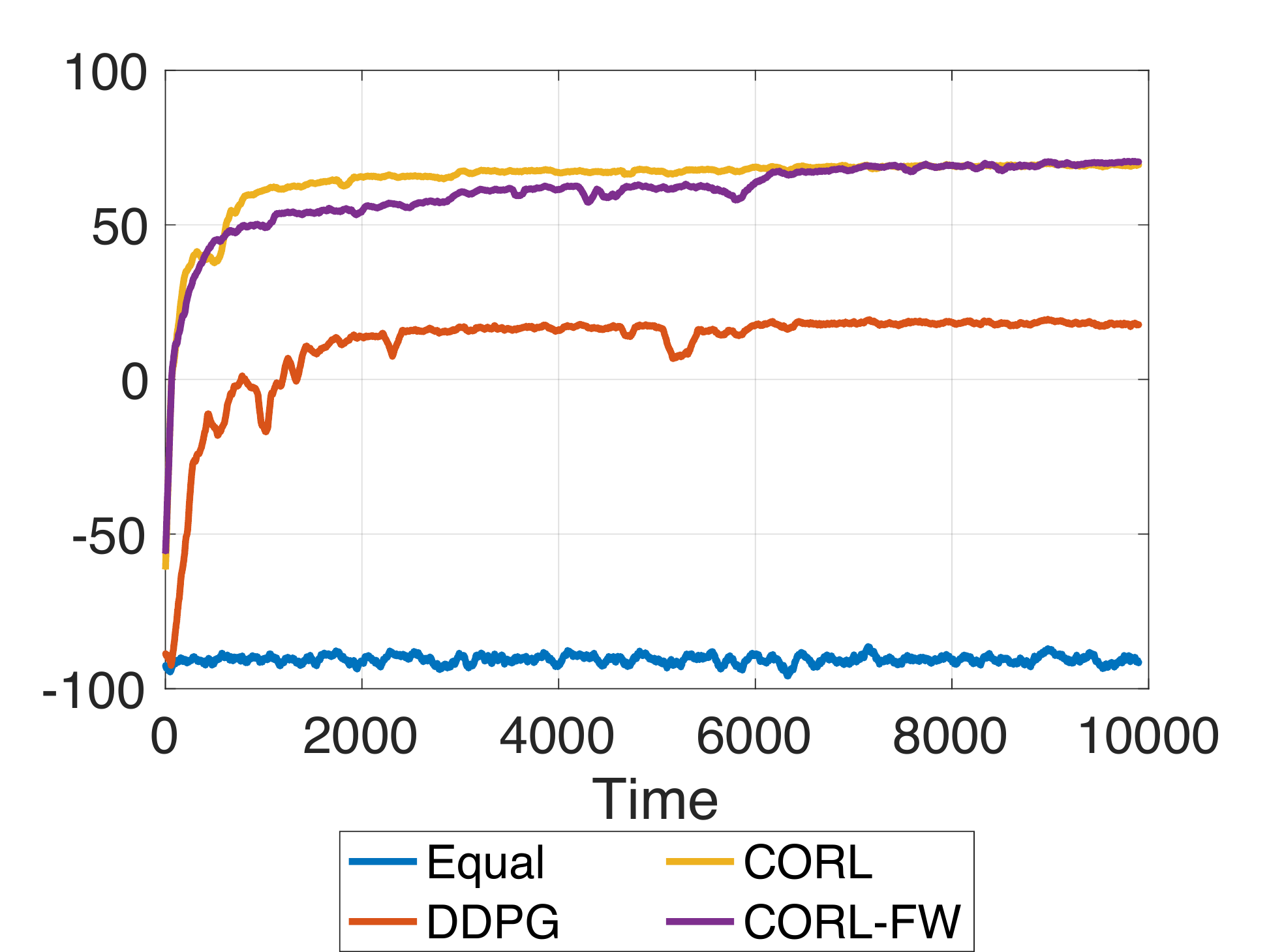}}%
\hfill
\subcaptionbox{rf6461}{%
\includegraphics[width=0.198\textwidth]{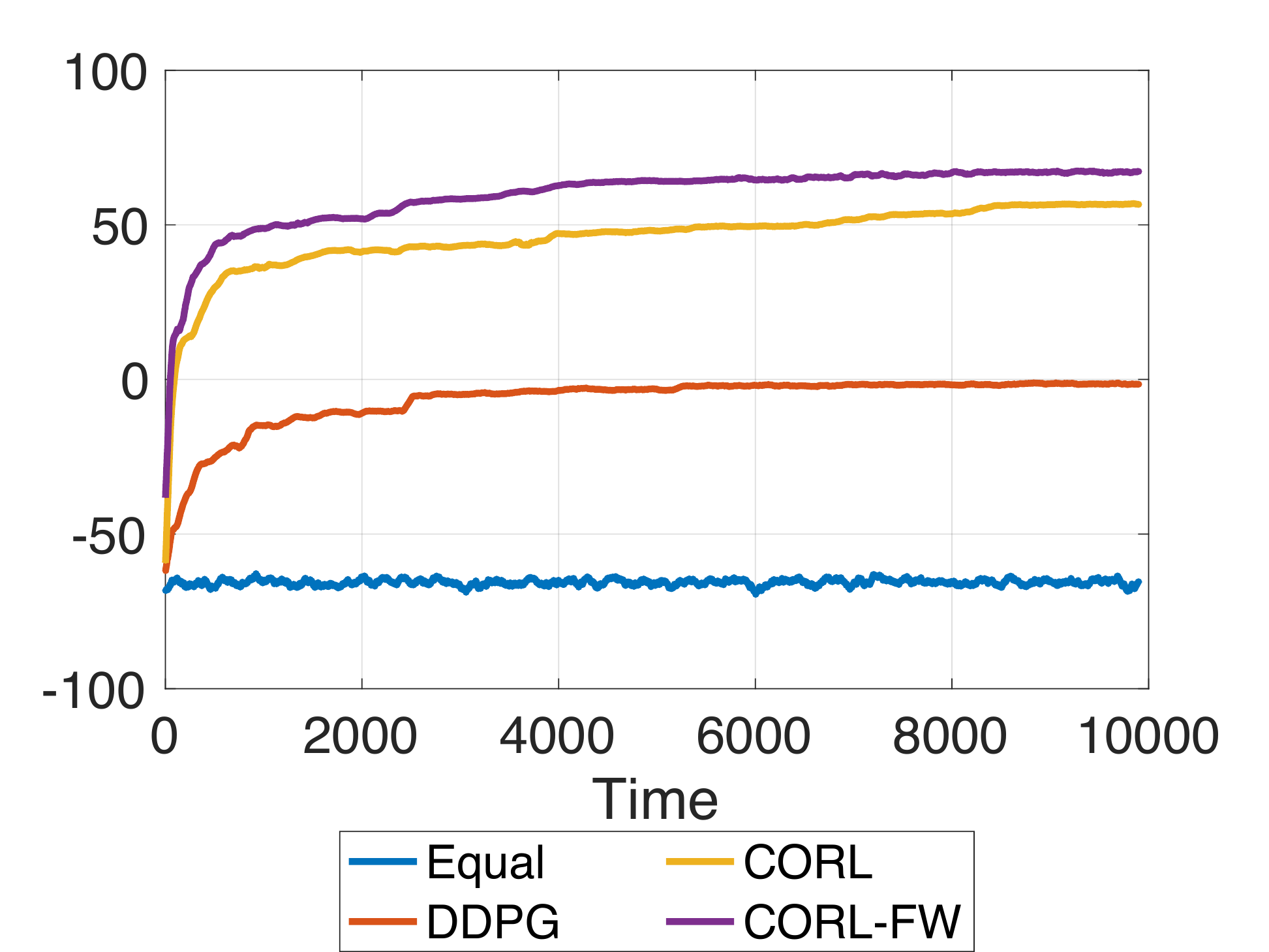}}%
\caption{Segement Routing}
\label{Fig:rf_SR}
\end{figure*}

For segment routing, after generating the mean values for each $p^{in}_i$ and $p^{out}_j$, we scale them up so that the maximum link utilization is over 105 percent (to show the need for avoiding highly utilized links). Similar to egress picking, for each topology 10 experiments are performed, with 4 randomly selected source nodes and 16 destinations. For the middle points, we include the 4 source nodes with an option of routing the traffic with no middle point and randomly select 12 other nodes as possible middle nodes. Results are shown in Figure \ref{Fig:rf_SR}. In this case, since the congested link generates a fixed delay providing no gradient, the direct FW approach fails to work. 

Since the congested link has to be avoided by spreading the traffic among certain paths, the problem of segment routing is harder than egress picking. For segment routing, the CORL methods achieve over 10 percent average delay reduction on rf1221 and 50 percent on the other four topologies, while DDPG fails to improve performance on four of the topologies. 

For segment routing, we also simulate the scenario where four service providers are simultaneously running the RL methods for congestion avoidance. Each service provider uses 4 source nodes, 16 middle points and 16 prefixes. 

\begin{figure*}[ht]
\subcaptionbox{rf1221}{%
\includegraphics[width=0.208\textwidth]{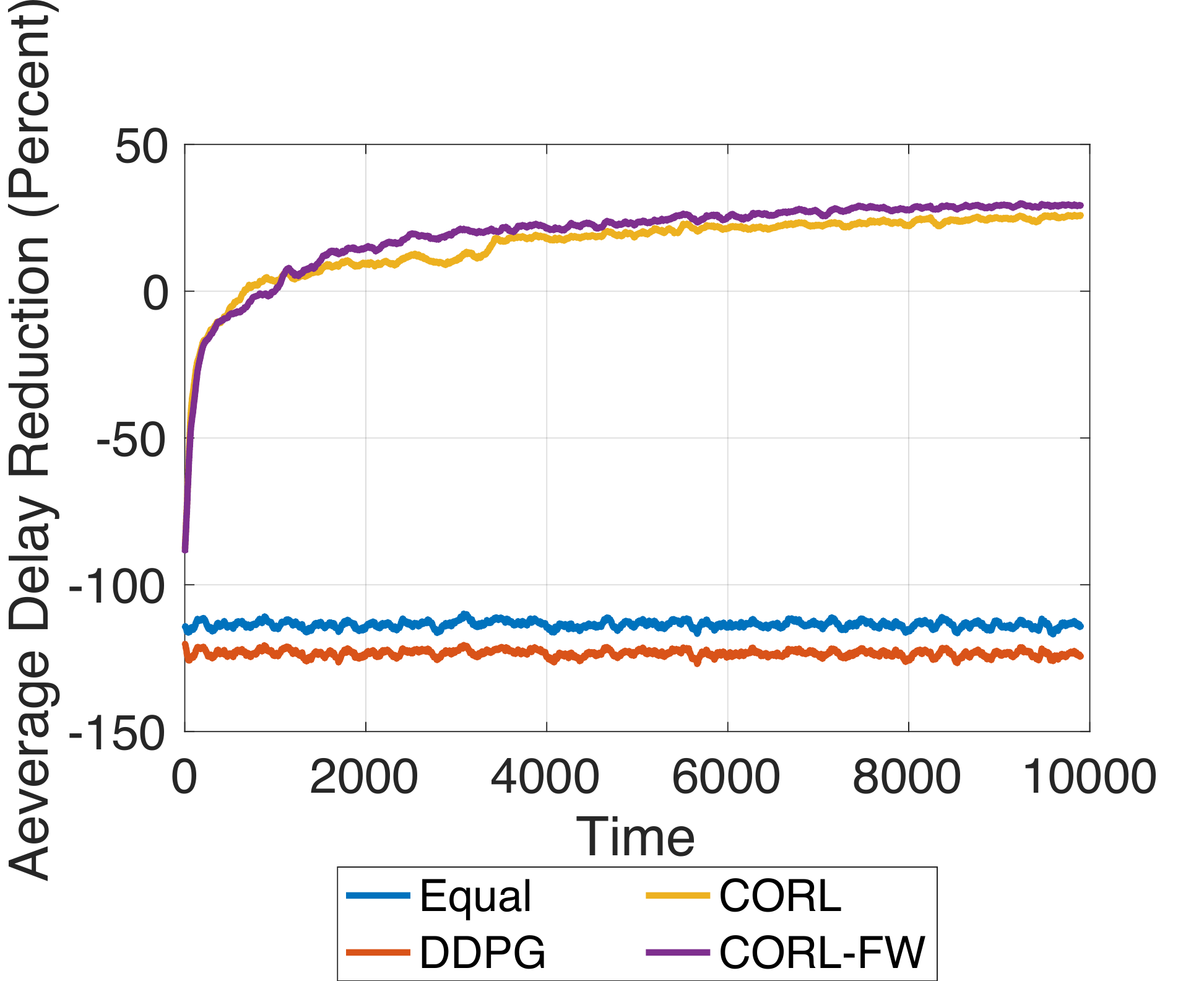}}%
\hfill
\subcaptionbox{rf1755}{%
\includegraphics[width=0.198\textwidth]{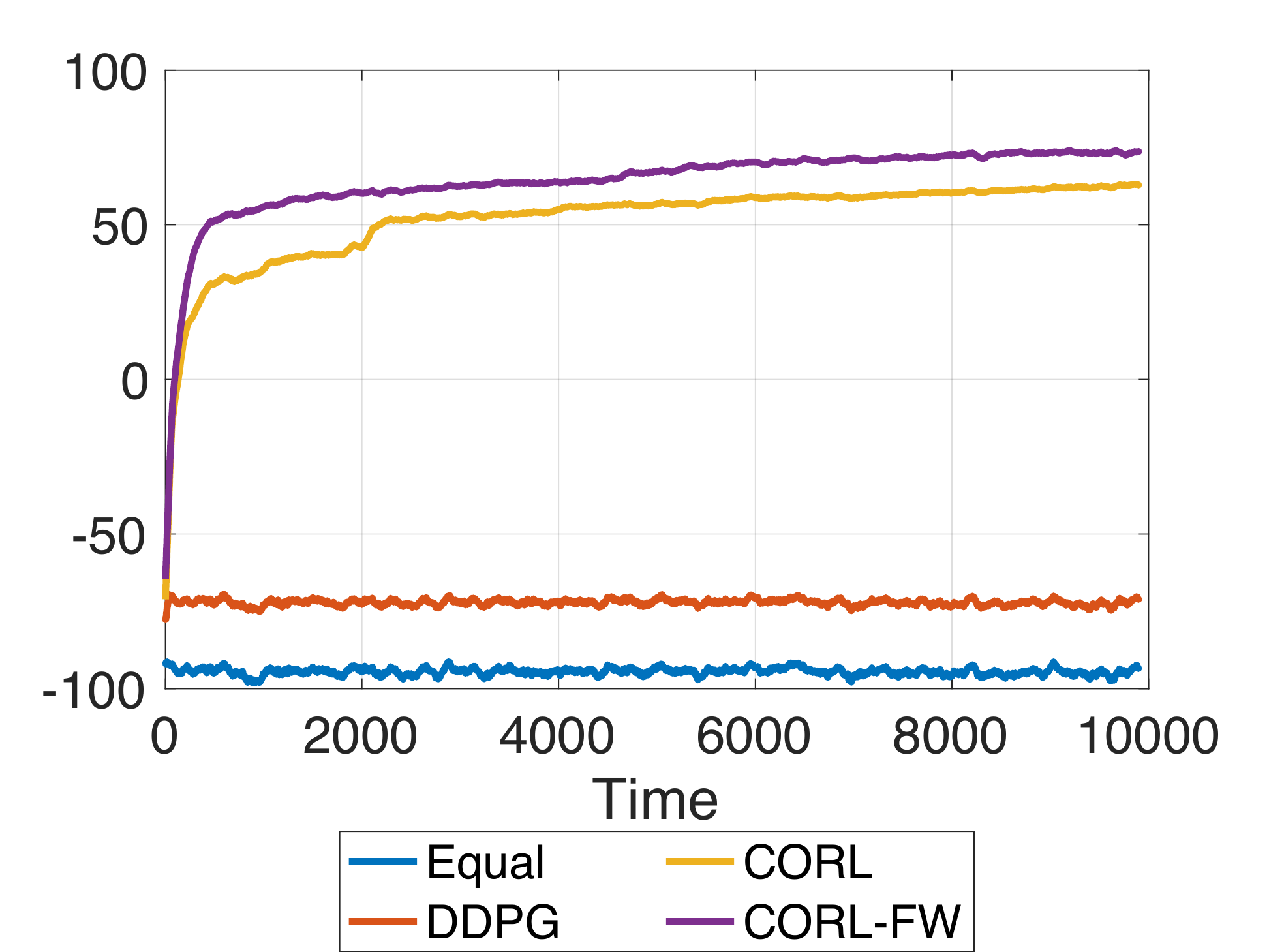}}%
\hfill
\subcaptionbox{rf3257}{%
\includegraphics[width=0.198\textwidth]{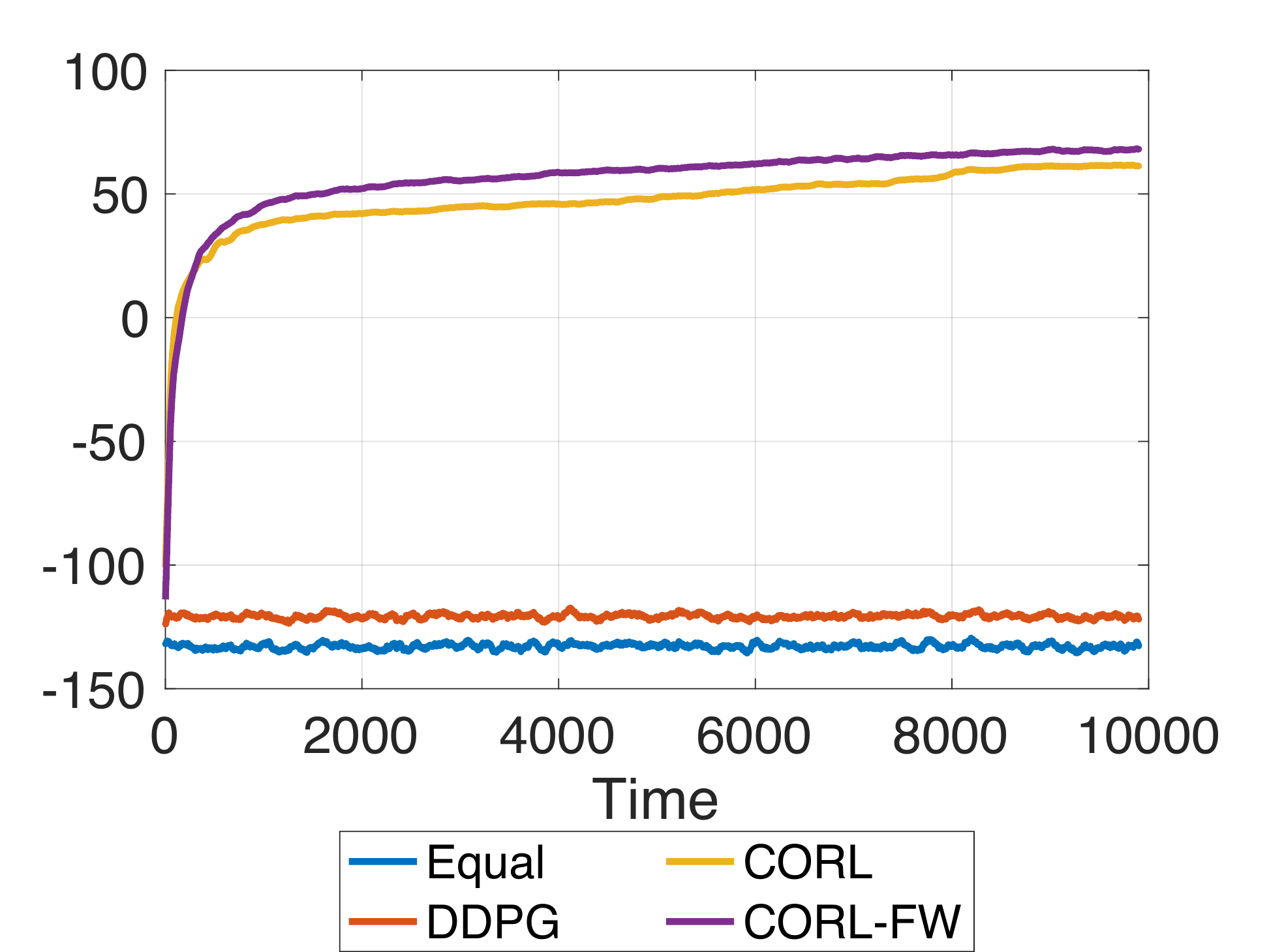}}%
\hfill
\subcaptionbox{rf3967}{%
\includegraphics[width=0.198\textwidth]{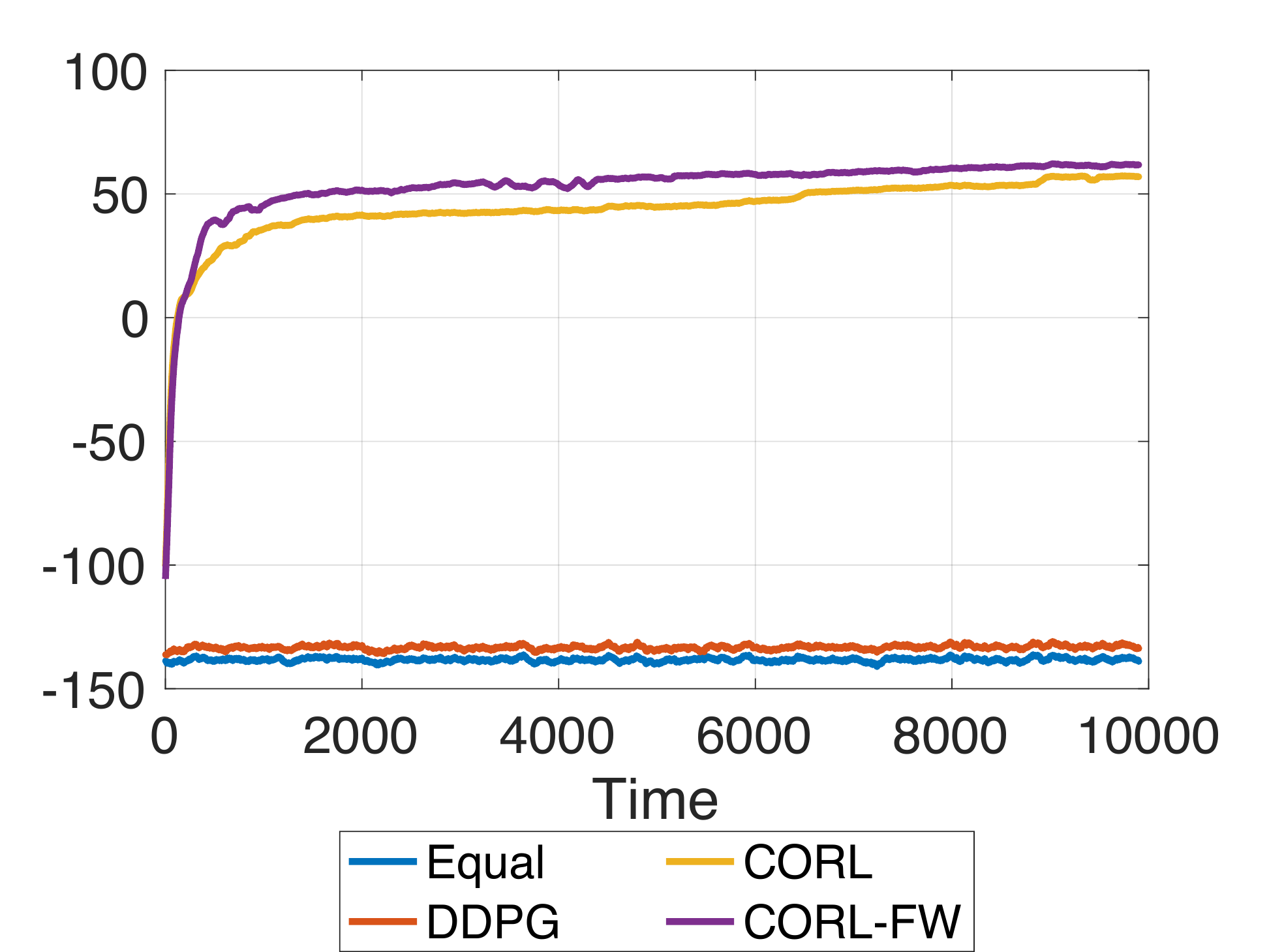}}%
\hfill
\subcaptionbox{rf6461}{%
\includegraphics[width=0.198\textwidth]{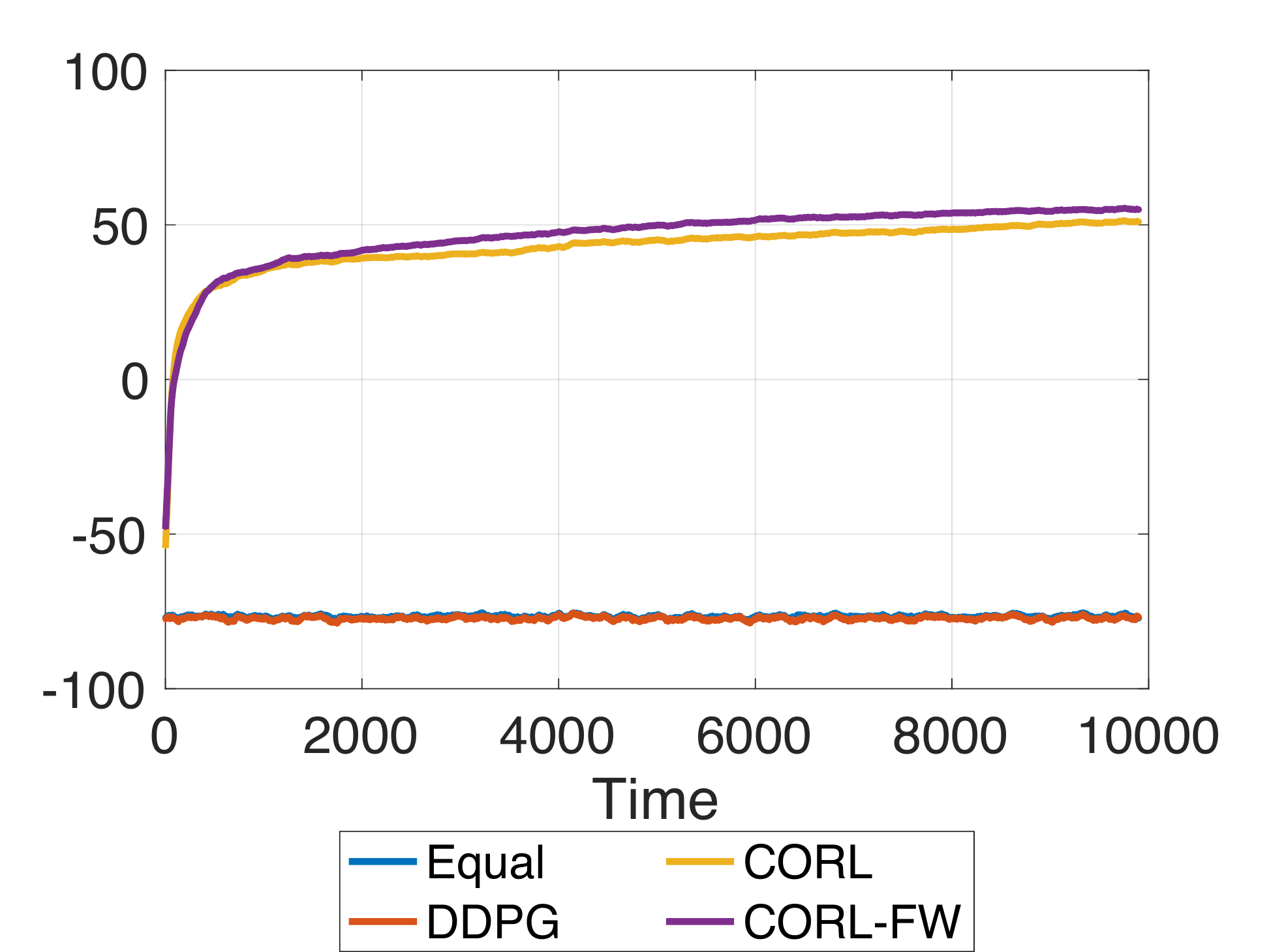}}%
\caption{Distributed Segment Routing}
\label{Fig:rf_SR_d}
\end{figure*}
In the distributed segment routing case, as shown in Figure \ref{Fig:rf_SR_d}, the CORL methods are still able to improve overall performance across the system, while DDPG and equal splitting fails to work in this case. Note that the DDPG is included only for comparison purposes since it is a widely used method. We do not advocate its use for our purpose. Clearly, traffic engineering using segment routing is much harder with only externally observable information. Nevertheless, we can still use learning-based methods to optimize performance as the experiments show. 

\subsection{Results of experiments on the Abilene dataset}
To further validate the performance of our methods with real traffic demands, we test our methods on the Abilene dataset \cite{zhang2003fast}. The Abilene dataset consists of about 40,0000 measurements of network traffic matrices (TM) on a topology with 12 nodes. The capacities of all the links in this topology are known, however the propagation delays are unknown. We use the known geographical locations to calculate the propagation delays $p$. We assume that we have control over four of the egress points (or associated ingresses into the black box network) in the topology.
The destinations includes all the other eight nodes. Each of the TMs is an average of traffic demands over 5 minutes. So the results on the Abilene dataset shows the performance of the methods over a longer range of time, with less frequent opportunities for changing the traffic splits. To simulate a realistic application of our methods, the TMs are replayed in time order. For each TM the learning agent is able to perform one splitting decision and obtain the delays between all four egress points and eight destinations. We ran ten experiments and each time the egress points and prefixes are selected randomly. 

\begin{figure}[t]
\centering
\includegraphics[width=.39\textwidth]{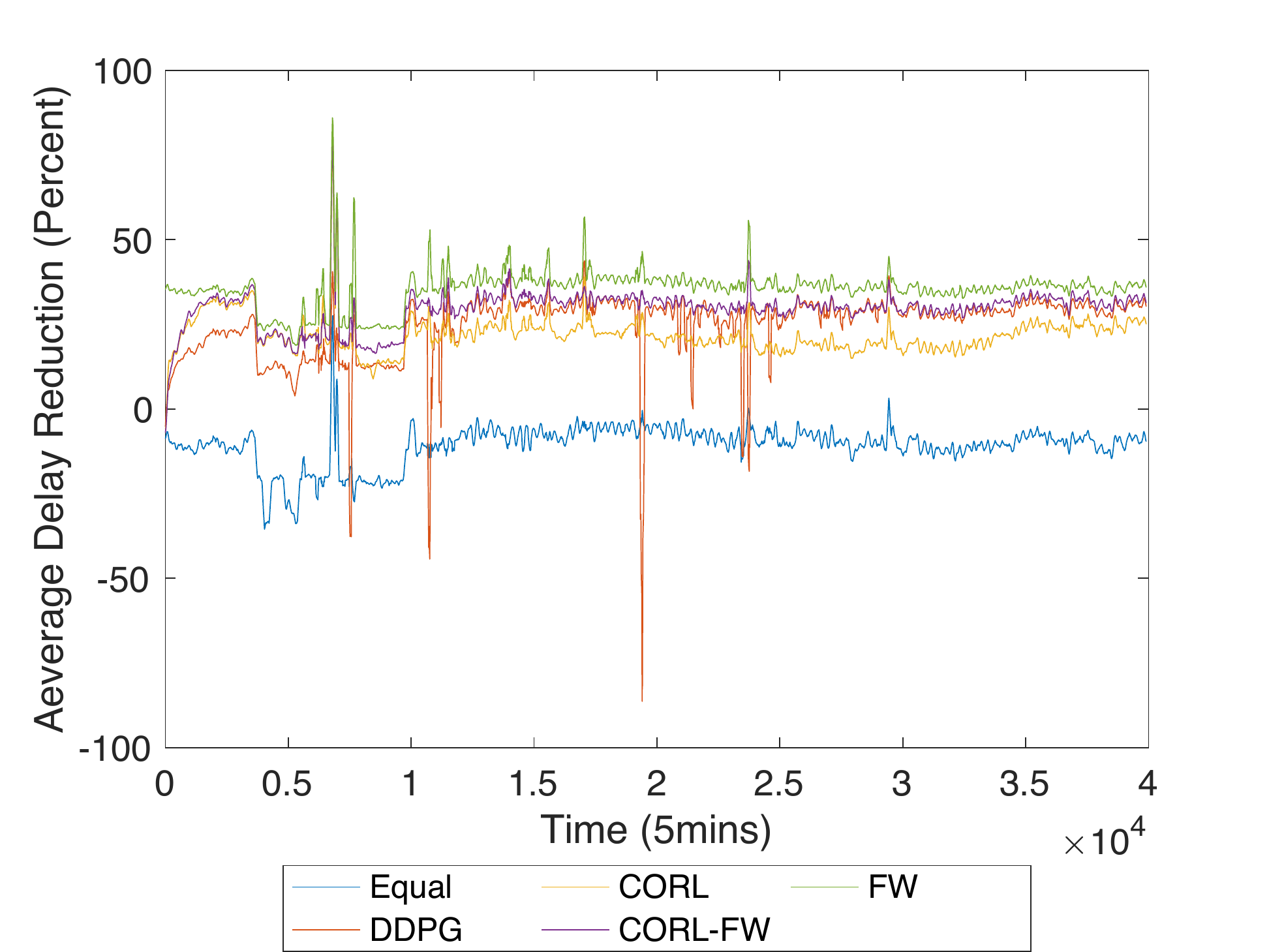}
\caption{Performance Comparison on Abilene Dataset}
\label{Fig: Abilene}
\end{figure}

As for the Rocketfuel topologies, for comparison, we also test the performance of DDPG \cite{lillicrap2015continuous}. For DDPG, CORL and CORL-FW we use NNs with two hidden layers of size 256. For all methods a replay buffer with size $1000$ and learning batch size of 32 is used. Soft update is adopted for better performance stability \cite{fox2015taming}. Figure \ref{Fig: Abilene} shows the performance of the two learning methods on the Abilene dataset. The delays are moving averages of 100 samples. For CORL and CORL-FW each action selection step consists of 100 iterations of gradient descent on the actions space, using the Adam optimizer \cite{kingma2014adam}. It can be seen that the Critic-Only method achieves lower delay with more robust performance compared with DDPG. 
To show the effectiveness of CORL methods, we also run a direct FW simulation. In this case we assume all the information about the link characteristics and other traffic are known. A step size of $\frac{2}{2+k}$ is used and we set the stopping criteria to be either reaching 100 steps of optimization or the distance between $\bm{D}$ and $\bm{x}_k$ is under 0.00001. The FW method converges before 100 iterations for over 90 percent of the cases. So it serves as a very close estimate of the lower bound of the average delay.

\subsection{Impact of Link Failures}
\begin{figure}[t]
	\centering
	\includegraphics[width=.39\textwidth]{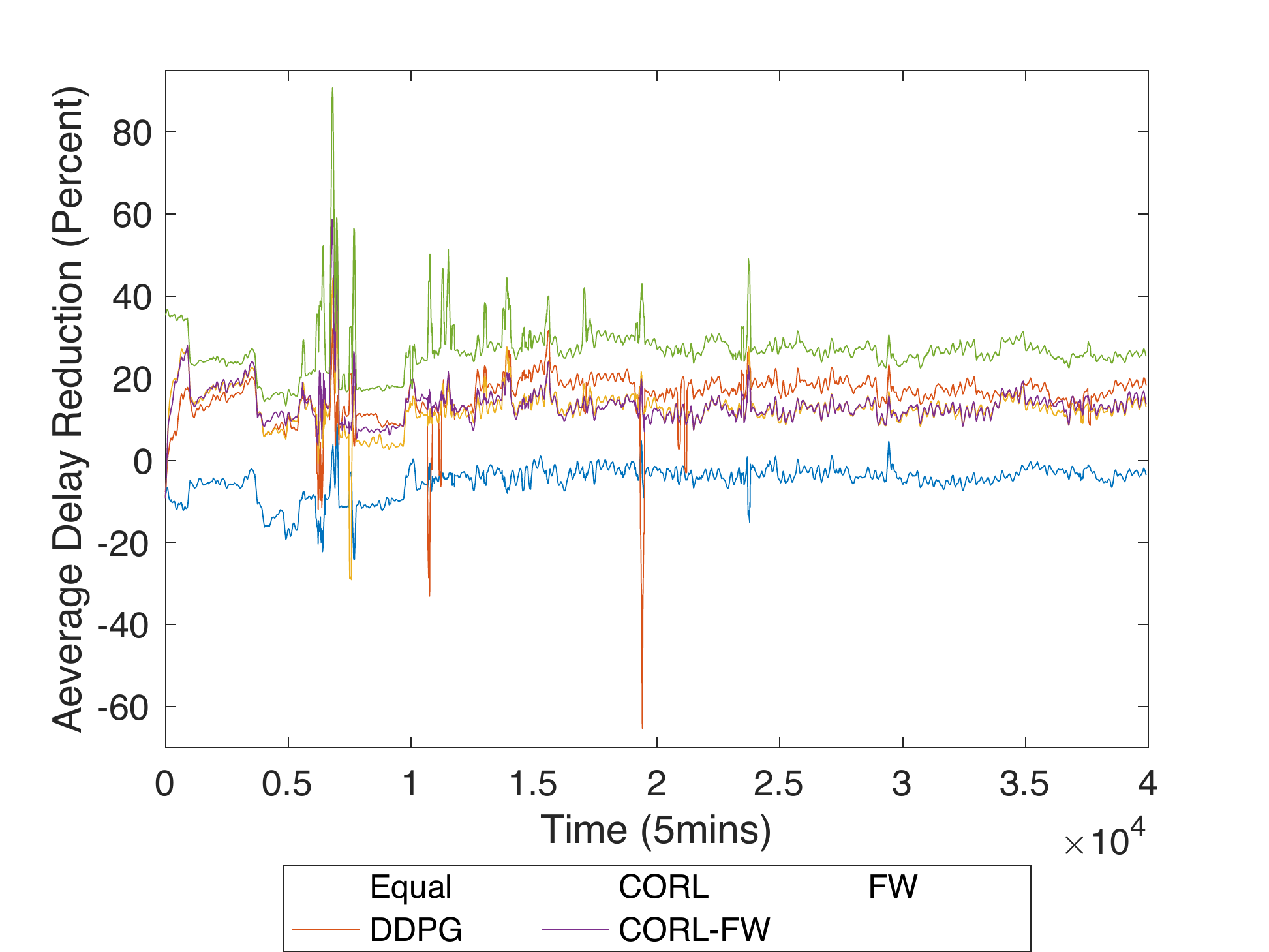}
\caption{Performance Comparison on Abilene Topology with Link Failure}
\label{Fig: Abilene_LF}
\end{figure}

We study the impact of link failures in the black box network. Though the black box network will have its own restoration mechanisms (such as IP or MPLS fast re-route) to handle link failures, clearly link failures can result in loss of network capacity and consequent delay increases. When external probing shows a delay increase, our decision making algorithms react to mitigate the delay increase. We perform experiments to study how well our algorithms respond to changes in network conditions. 
To study the effect of link failure, for the first 1000 samples the original topology is used. Then after every 1000 samples an impaired topology is used. The topology is generated by randomly dropping one link from the original graph, while still keeping all the nodes connected. Results are shown in Figure \ref{Fig: Abilene_LF}. In this case CORL-FW achieves a good compromise between close to optimum results and robust performance, showing more stable performance after link failures.

\section{Concluding Remarks}
Network tomography has been extensively studied as a means to infer pertinent network characteristics from external observations. With the growing use of network virtualization and deployment of overlay technologies such as SD-WANs, the use of network tomography by overlay networks to infer characteristics of the underlay networks is likely to grow. In this paper, we considered the use of network tomography in a broader context -- can we combine network tomography with machine learning based decision making for automated performance optimization from the periphery of the network. We considered this problem in a general setting where an external entity that is generating traffic (such as an over the top service provider) to a set of its clients (destinations) has to send the traffic over a black box network (such as an underlay) while minimizing average delay. Using learning-based approaches for this problem poses several technical challenges including the dimension of the solution space and the enforcement of constraints that arise naturally in the networking context (such as policy constraints, capacity constraints, etc.). We show how these constraints can be handled while using a deep reinforcement learning framework for decision making. For two representative problems (egress traffic picking and optimized segment routing) we show experimental results on two widely used network topology databases. The methods we use can be used both in a centralized manner and distributedly by multiple independent agents. The effectiveness of our method is illustrated by delay reductions of as much as 60\% in comparison to standard heuristics.  We believe that the use of tomography with machine learning has many other applications such as in wireless networks.

\balance

\bibliographystyle{IEEEtran}
\bibliography{reference}

\end{document}